\documentclass[manuscript,screen,acmsmall]{acmart}
\AtBeginDocument{%
  }

\usepackage{booktabs}   %% For formal tables:
\usepackage{subcaption} %% For complex figures with subfigures/subcaptions
\usepackage{algorithm}
\usepackage{multirow}
\usepackage[noend]{algorithmic}

\usepackage{courier} %% Sets font for listing as Courier.
\usepackage{xspace}
\usepackage{framed}
\usepackage{listings, xcolor}
\usepackage{xcolor}
\usepackage{soul}
\usepackage{caption}

\usepackage{tcolorbox}
\tcbuselibrary{listingsutf8}
\tcbuselibrary{breakable, listings}

\usepackage{booktabs}
\definecolor{colortcli}{HTML}{FF6699}
\def\toolname{\textsc{LORIS}\xspace{}}
\def\baselinename{\textsc{LaM4Inv}\xspace{}}
\def\clausetoinv{\textsc{Clause2Inv}\xspace{}}

\newtcblisting{PromptBox}{
    colback=gray!10!white,
    colframe=blue!75!black,
    listing only,
    left=2mm, right=2mm, top=1mm, bottom=1mm,
    title={Prompt},
    boxrule=0.7pt,
    arc=2mm,
    listing options={
        basicstyle=\ttfamily\small,
        breaklines=true,
        breakindent=0pt,
        numbers=none,
        language={},  % 普通文本
        tabsize=4,  % 设置制表符宽度
    }
}

\newtcblisting{BPromptBox}{
    breakable, % For page breaks
    colback=gray!10!white,
    colframe=blue!75!black,
    listing only,
    left=2mm, right=2mm, top=1mm, bottom=1mm,
    title={Prompt},
    boxrule=0.7pt,
    arc=2mm,
    listing options={
        basicstyle=\ttfamily\small,
        breaklines=true,
        breakindent=0pt,
        numbers=none,
        language={},  % 普通文本
        tabsize=4,  % 设置制表符宽度
    }
}

\def\imply{\Rightarrow}

\definecolor{colorzyyan}{HTML}{F6a6a6}

\begin{document}

%% Title information
\title{Guiding LLM-based Loop Invariant Synthesis via Feedback on Local Reasoning Errors }         %% [Short Title] is optional;
                                        %% when present, will be used in
                                        %% header instead of Full Title.
                                        %% can be repeated if necessary;
                                        %% contents suppressed with 'anonymous'

%% Author information
%% Contents and number of authors suppressed with 'anonymous'.
%% Each author should be introduced by \author, followed by
%% \authornote (optional), \orcid (optional), \affiliation, and
%% \email.
%% An author may have multiple affiliations and/or emails; repeat the
%% appropriate command.
%% Many elements are not rendered, but should be provided for metadata
%% extraction tools.

\author{Tianchi Li}
\affiliation{
  \department{Key Laboratory of High Confidence Software Technologies (Peking University), Ministry of Education; School of Computer Science}              %% \department is recommended
  \institution{Peking University}            %% \institution is required
  \city{Beijing}
  \country{China}                    %% \country is recommended
}
\email{litianchi@pku.edu.cn}  
\orcid{0009-0009-4593-2737}
%% Author with single affiliation.

\author{Zhenyu Yan}
% \thanks{These authors contributed equally to this work.}
\authornote{These authors contributed equally to this work.}
\orcid{0009-0007-6181-6205}
\affiliation{%
  \department{Key Laboratory of High Confidence Software Technologies (Peking University), Ministry of Education; School of Computer Science}              %% \department is recommended
  \institution{Peking University}
  \country{China}
}
\email{zhenyuyan@stu.pku.edu.cn}          %% \email is recommended

\author{Junhao Liu}
\authornotemark[1]
% \authornote{These authors contributed equally to this work.}
% \thanks{These authors contributed equally to this work.}
\orcid{0009-0004-9949-6583}
\affiliation{%
  \department{Key Laboratory of High Confidence Software Technologies (Peking University), Ministry of Education; School of Computer Science}              %% \department is recommended
  \institution{Peking University}
  \country{China}
}
\email{liujunhao@pku.edu.cn}

\author{Peng Di}
\orcid{https://orcid.org/0000-0002-5799-5876}             %% \orcid is optional
\affiliation{%
  \institution{Ant Group}
  \city{Hangzhou}
  \state{Zhejiang}
  \country{China}
}
\email{dipeng.dp@antgroup.com}

\author{Xin Zhang}
\authornote{Corresponding author.}
\affiliation{
  \department{Key Laboratory of High Confidence Software Technologies (Peking University), Ministry of Education; School of Computer Science}              %% \department is recommended
  \institution{Peking University}            %% \institution is required
  \city{Beijing}
  \country{China}                    %% \country is recommended
}
\email{xin@pku.edu.cn}          %% \email is recommended
\orcid{0000-0002-1515-7145}
%% Abstract
%% Note: \begin{abstract}...\end{abstract} environment must come
%% before \maketitle command
\begin{abstract}

%The ``guess-and-check'' paradigm is a powerful problem-solving strategy, where Large Language Models (LLMs) have shown remarkable promise in the ``guess'' step, particularly for tasks like program verification.
%
%However, if an LLM-generated solution is incorrect, the formal ``checker'' typically provides binary pass/fail feedback, which is insufficient to guide the LLM toward a correct answer.
%
%To address this,
We propose a novel framework that provides constructive feedback to an LLM in the ``guess-and-check'' paradigm by formally verifying its own thinking process and detecting local reasoning errors.
We apply this framework to the loop invariant synthesis problem.
We prompt the model to produce a step-by-step natural language proof justifying its thinking process for the failed verification condition of its generated loop invariants.
%When an LLM-proposed invariant fails verification, we prompt the model to produce a step-by-step natural language proof justifying its thinking process for the specific failed verification condition.
%
%We introduce a lightweight auto-formalization technique to formalize its thinking process.
%
Then, we use an LLM to translate the reasoning steps into first-order logic implications, which can be checked automatically.
An invalid implication pinpoints the exact logical flaw in the LLM's thinking process, which we then use to construct targeted feedback for refinement.
We have implemented our approach in a tool called \toolname{} and evaluated it on a main benchmark suite of 460 C programs and an additional benchmark suite of 50 C programs each of which involves non-linear properties.
On the main benchmark suite, LORIS solved 445 of the programs, and achieved an overall success rate of 93.1\%.
LORIS also demonstrates robustness on the challenging non-linear benchmark suite.
%We evaluated \toolname{} on a benchmark of 460 C programs requiring loop invariants. Our results show that by leveraging our feedback mechanism, \toolname{} solves 445 of the problems, demonstrating that formally checking the LLM's local reasoning is a highly effective strategy for guiding it to verifiably correct solutions.

\end{abstract}

%% 2012 ACM Computing Classification System (CSS) concepts
%% Generate at 'http://dl.acm.org/ccs/ccs.cfm'.
%\begin{CCSXML}
%<ccs2012>
%<concept>
%<concept_id>10011007.10011006.10011008</concept_id>
%<concept_desc>Software and its engineering~General programming languages</concept_desc>
%<concept_significance>500</concept_significance>
%</concept>
%<concept>
%<concept_id>10003456.10003457.10003521.10003525</concept_id>
%<concept_desc>Social and professional topics~History of programming languages</concept_desc>
%<concept_significance>300</concept_significance>
%</concept>
%</ccs2012>
%\end{CCSXML}

%\ccsdesc[500]{Software and its engineering~Automated Static Analysis}
%\ccsdesc[300]{Social and professional topics~History of programming languages}
%% End of generated code


%% Keywords
%% comma separated list
  %% \keywords are mandatory in final camera-ready submission

%% \maketitle
%% Note: \maketitle command must come after title commands, author
%% commands, abstract environment, Computing Classification System
%% environment and commands, and keywords command.
\setcopyright{cc}
\setcctype{by}
\acmJournal{TOPLAS}
\acmYear{2026} \acmVolume{48} \acmNumber{2} \acmArticle{8}
\acmMonth{5} \acmDOI{10.1145/3806652}

\maketitle

\section{Introduction}
\label{sec:intro}

The ``guess-and-check'' paradigm is a fundamental problem-solving strategy across various domains, such as mathematical theorem proving, program analysis, program verification, and so on.
In this paradigm, a potential solution is first proposed (the ``guess'' step), and then the correctness of the solution is verified (the ``check'' step).
%
% GPT (generative pre-trained transformer) is the name of a type of LLMs, a specific LLM name here maybe better
% OLD: Recently, Large Language Models (LLMs), such as GPT, have demonstrated their remarkable capabilities in the ``guess'' phase.
Recently, Large Language Models (LLMs), such as GPT series~\cite{ChatGPT4} and DeepSeek series~\cite{DeepSeekV3, DeepSeekR1}, have emerged as promising approaches to solve the ``guess'' step. 
% \yannew{in program understanding and generation}~\cite{FangMSLZ24,jiang2024surveylargelanguagemodels}.
% \todotcli{Why mentioning program-related tasks here?}
%
With their extensive training on huge amounts of data, LLMs can generate creative and high-quality candidate solutions, which often capture the high-level intuition required to solve complex problems.
Moreover, while LLMs have no correctness guarantees and can generate erroneous solutions~\cite{HuangYMZFZQW25,surveylogicerror}, the ``check'' step in the paradigm can reject incorrect proposals and therefore address this shortcoming, making the ``guess-and-check'' paradigm particularly suitable for applying LLMs.
However, one challenge is that when LLMs fail to generate correct solutions, current ``checkers'' cannot provide fine-grained feedback to guide LLMs to improve their outputs.
This paper tries to address this problem by generating constructive feedback via formally checking the thinking process of an LLM.
In particular, we focus on subtle logical errors as LLMs often think in a generally right direction but make mistakes in the details due to issues like \emph{hallucination}~\cite{HuangYMZFZQW25}.
% with loop invariant synthesis as a representative application. 
% However, a significant limitation of current LLMs is the lack of correctness guarantees in their outputs: issues such as \emph{hallucination}~\cite{HuangYMZFZQW25} and \emph{logical errors}~\cite{surveylogicerror} are prevalent in LLM-generated solutions.
%
% OLD: This presents a promising opportunity: by systematically identifying and feeding back such detailed errors to the LLM, LLM can be guided to refine its initial high-level concept into a verifiably correct solution.
%This \yannew{raises a challenge of how to safely apply LLMs to solve those questions but also} presents a promising opportunity: by systematically identifying and feeding back such detailed errors to the LLM, LLM can be guided to refine its initial high-level concept into a verifiably correct solution.
% This presents a promising opportunity: Instead of demanding LLMs to produce rigorous, verifiably correct solutions, we can leverage them to guess the answer within the guess-and-check framework.
% The core idea is to pair a creative LLM ``guesser'' with a formal ``checker''.
% To help the LLM  guess a correct answer, we can provide systematic and detailed feedback on the thinking process of why it thinks its answer is correct.

%
% While the overall strategy of an LLM-generated solution might be correct, it often contains subtle but critical errors in the details.
%
% The hallucination can take various forms, such as incorrect logical deductions or the use of inaccurate conditions.
%
%

%
Loop invariant synthesis, a cornerstone task in program analysis and verification, is a good example to explore this feedback-guided approach.
A loop invariant is a property that holds before and after each iteration of a loop. Identifying correct and sufficiently strong invariants is essential for formally proving program correctness.
Over the past decades, numerous techniques have been developed to automate loop invariant synthesis, most of which follow the ``guess-and-check'' paradigm~\cite{SomenziB11,ErnstCGN00,GabrielJJYJ19,GargLMDN14,SiNDNS20}.
In this paradigm, a candidate invariant is first generated, and then the correctness of the invariant and the program property is checked by a verification tool.
If the invariant fails to pass the check, certain feedback, such as counterexamples, will be generated to guide the approach to find another invariant until the correct one is generated.
Existing techniques for generating candidate loop invariants include symbolic and data-driven approaches.
% Remove Xujie's SiNDNS20 as it allows running online RL with an uninited model.
These approaches either rely on the pre-defined templates~\cite{SomenziB11,ErnstCGN00}, or require a large amount of training data~\cite{GabrielJJYJ19,GargLMDN14}.
On the other hand, although pre-trained LLMs have shown their abilities in comprehending and reasoning about programs, they still struggle to generate correct and complete loop invariants directly~\cite{AdharshKSLDRR23,PeiBSY23}.
To address this, we propose a novel framework that leverages LLMs to synthesize loop invariants, guided by formal verification feedback on the model's own thinking process.
In our approach, an LLM first proposes a candidate loop invariant for a given program and property.
If a verifier confirms that the invariant is both correct and sufficient to prove the program property, the synthesis process terminates successfully.
Otherwise, we prompt the LLM to produce a step-by-step thinking process to prove the correctness of its proposed invariant.
We then formally analyze its thinking process to detect exact logical errors and provide them as precise feedback to the LLM for refinement.
This process iterates until a correct invariant is generated.
%

% However, there are some significant challenges.
% %
% First, the proofs of program properties using loop invariants can be long and complex, making it difficult for LLMs to generate a complete step-by-step proof in a single response.
% %
% Second, the LLM's thinking process is typically expressed in natural language, which raises the question of how to formally and automatically detect errors within this unstructured text.
% %
% Third, the feedback to the LLM should be constructive. It should be as precise as possible in reflecting the logical errors in the model's thinking process.
%

The key challenges in our framework are how to formalize the thinking process of an LLM and how to provide constructive feedback based on the verification result of the thinking process.
To address the challenges, we leverage the idea of autoformalization~\cite{WuJLRSJS22}, where the LLM generates its thinking process in natural languages, and an LLM (potentially a different one) translates the thinking process into a formal proof.
However, in contrast to existing approaches~\cite{jiang2023draft,zhou2024dont} that produce fully checkable proof scripts in formal systems such as Coq or Isabelle, we intentionally avoid this strategy due to the following considerations.
First, proofs in these formal languages are quite different from their natural language counterparts, making the translation potentially unfaithful.
Second, LLMs often generate incorrect proofs due to low-level tactic issues, such as incorrect theorem applications~\cite{LuDZ24}.
Third, it is difficult to map the failure of the proof script back to the natural language thinking process, thereby preventing effective feedback.

To address the above issues, our key observation is that LLMs often think in generally right directions but fail in the details. For example, it is very common for LLMs to have off-by-one errors.
Following this observation, instead of checking the full correctness of the thinking process, we focus on checking \emph{local correctness} to provide effective feedback to LLMs on subtle errors.
Specifically, we instruct the LLM to formalize the key steps in the natural language proof into a series of first-order logic implications.
In this way, we do not check the correctness of the high-level strategy but focus on the correctness of each individual step.
These implications are lightweight and can be efficiently checked for validity by an SMT solver.
An invalid implication discovered by the SMT solver corresponds directly to a specific logical error in the LLM's thinking process, which can form constructive feedback to the LLM.

We have implemented our framework as a tool called \toolname{} (\textbf{LO}cal \textbf{R}easoning-guided \textbf{I}nvariant \textbf{S}ynthesizer), which synthesizes loop invariants to verify properties of C programs.
We evaluated \toolname{} on a main benchmark suite of $460$ C programs requiring loop invariants for verification, and an additional benchmark suite of 50 C programs each of which involves  non-linear properties.
Our experiment results show that by leveraging GPT-4.1, our tool can successfully solve $445$ of the programs, reaching an overall success rate of 93.1\% on the main benchmark suite.
On the challenging non-linear benchmark suite, our tool solves 47 of the 50 programs.
These results show that the feedback framework provided by our approach can guide the LLMs toward correct solutions more effectively.
% OLD: Furthermore, when using the reasoning model GPT-o4-mini,
%\yannew{To demonstrate the generality of our approach, we conducts our experiments on another model}, GPT-o4-mini. \yannew{On GPT-o4-mini}, Our tool achieves a success rate $14.9\%$ higher than the baseline approach,\TODO{[Provide baseline's rate (from $xx\%$ to $yy\%$)]} showing that the feedback \yannew{framework} provided by our approach can guide the \yannew{LLMs} toward correct solutions more effectively.
%

%
In summary, our contributions are as follows:
\begin{enumerate}
    \item We propose a novel paradigm that guides an LLM to refine its solution within the ``guess-and-check'' paradigm, by providing formal verification feedback on the model's own thinking process.
    \item We propose a framework for the loop invariant synthesis using our proposed paradigm, where an LLM's natural language reasoning is formalized into first-order logic implications and checked by an SMT solver to generate precise feedback. We have implemented our framework as a tool called \toolname{}.
    \item We have demonstrated the effectiveness of our approach on a main benchmark suite of $460$ C programs requiring loop invariants for verification, and an additional benchmark suite of $50$ C programs each of which involves non-linear properties. The result shows that our approach effectively improves the number of correct solutions proposed by LLMs.
    
\end{enumerate}

\section{Motivating Example}
\label{sec:example}

\begin{figure}[t]
    \centering

    \begin{minipage}{0.44\textwidth}

    \begin{lstlisting}[language=C,basicstyle=\ttfamily\small]
extern int unknown();

int main() {
    int a = 0;
    int j, m;
    if(m <= 0) return 0;

    for(j = 1; j <= m; j++) {
        if(unknown())
            a++;
        else 
            a--;
    }
    
    assert(a >= -m && a <= m);
    return 0;
}
    \end{lstlisting}
    \caption{Example program.}
    \label{fig:example-program}

    \end{minipage}
\hfill
    \begin{minipage}{0.52\textwidth}
    \begin{minipage}{\linewidth}
    \begin{lstlisting}[language=C,basicstyle=\ttfamily\small]
/*@
    loop invariant i1: a >= -(j - 1) && a <= (j - 1);
    loop invariant i2: j >= 1;
    loop invariant i3: m > 0;
*/
    \end{lstlisting}
    \caption{Loop invariants proposed by the LLM.}
    \vspace{0.2in}
    \label{fig:example-propose}
    \end{minipage}
%\vspace{2in}
    \begin{minipage}{\linewidth}
\begin{lstlisting}[language=C,basicstyle=\ttfamily\small]
/*@
    loop invariant i1: a >= -j + 1 && a <= j - 1;
    loop invariant i2: j >= 1;
    loop invariant i3: j <= m + 1;
    loop invariant i4: m > 0;
*/
    \end{lstlisting}
    \caption{Loop invariants refined by the LLM.}
    \label{fig:example-refine}
    \end{minipage}
    \end{minipage}

\end{figure}

In this section, we use a verification problem for a simple C program to illustrate our approach.
Figure~\ref{fig:example-program} shows a C program with a loop and an assertion.
Inside the loop, the variable $a$ is non-deterministically incremented or decremented by one, depending on the return value of the function \texttt{unknown}.
The loop iterates $m$ times, where $m$ is a positive integer.
The verification task is to prove that after the loop, the final value of $a$ is bounded by $m$, i.e., $-m \leq a \leq m$.
%

%
%In order to verify the assertion, we need to propose some loop invariants.
To prove the assertion, correct loop invariants must be discovered first.
A loop invariant is a program property that holds before the loop begins and is preserved by each iteration.
In this example, a sufficient set of loop invariants is: 1) $-(j-1)\leq a\leq (j-1)$ and 2) $ 1\leq j \leq (m+1)$.
%
%Then, we need to verify the correctness of the loop invariants, and use the proposed invariants to verify the assertion.
%
The verification of the program's assertion then relies on checking the following three conditions based on these invariants:

%\todotcli{Names like $Pre, Inv$ are not defined here. Is it proper to use them?}
\begin{enumerate}
    \item \textit{Establishment.} The loop invariants hold before the loop starts ($Pre\Rightarrow Inv$). The program's pre-conditions must imply the invariants:
    \[
    (a=0\ \land\ j=1\ \land\ m>0)\implies (-(j-1)\leq a\leq (j-1)\ \land\ 1\leq j \leq (m+1))
    \]
    \item \textit{Preservation.} If the invariants hold at the beginning of an iteration, they must also hold at the end ($\{Inv\wedge Cond\}Prog\{Inv\}$). We use a Hoare triple to represent this condition, where $S$ is the loop body:
   \[
\begin{aligned}
\{\, & -(j-1)\le a\le (j-1)
       \;\land\; 1\le j\le m+1
       \;\land\; j\le m \,\}\; S \\[4pt]   % ← \\ 换行并稍微留一点竖向间距
\{\, & -(j-1)\le a\le (j-1)
       \;\land\; 1\le j\le m+1 \,\}
\end{aligned}
\]
    \item \textit{Post-condition.} When the loop terminates, the invariants must imply the desired program assertion ($Inv\wedge\neg Cond\Rightarrow Post$):
    \[
    (-(j-1)\leq a\leq (j-1)\ \land\ 1\leq j \leq (m+1)\ \land\ \lnot(j\leq m))\implies (-m\leq a\leq m)
    \]
\end{enumerate}

%In the preservation condition, we use a Hoare-triple to represent if the pre-condition holds after the loop body $B$, then the post-condition holds after the loop body.
%

%
% This paradigm does not apply to ALL invariant generation approaches, right?
The overall verification process of our approach follows the ``guess-and-check'' paradigm.
That is, candidate loop invariants are first generated in the ``guess'' step, and the three conditions above are then formally verified in the ``check'' step, which can be efficiently handled by modern verification tools.

%For the ``guess'' step, previous works have developed various techniques to find the correct loop invariants efficiently.
For the ``guess'' step, previous works have developed various techniques to find loop invariants, including template-based and learning-based methods\cite{antopoulosCounterexampleguidedApproachFinding2016}.
%
%These techniques include template-based methods, which propose parameterized loop invariants from pre-defined templates, involving some heuristic strategies.
%ß
%Learning-based methods are also used to generate candidate loop invariants, which synthesize invariants using a machine-learning model or a neural network.
%
%Recently, with the rapid development of Large Language Models (LLMs), they have shown strong performance in comprehending and reasoning about programs.
Recently, the advances in comprehending and reasoning about programs shown by Large Language Models (LLMs) suggest their potential for generating invariants directly from a zero-shot prompt.
%
%It is potential to let LLMs generate loop invariants directly in a zero-shot prompt.
%
%
Motivated by these insights, we prompt ChatGPT-4o to generate loop invariants to verify the assertion for the C program in Figure~\ref{fig:example-program}.
It responds with the candidates shown in Figure~\ref{fig:example-propose}.
As we can see, the loop invariant $i1$ correctly captures the relationship between $a$ and the loop counter $j$.
While the loop invariant $i2$ tries to identify the bound of $j$, it omits the upper bound, $j <= m+1$.
Although the proposed invariants are individually correct, they are insufficient to prove the assertion.
This is because, without the upper bound, the exact value of $j$ at the loop termination cannot be precisely determined from the loop exit condition $\lnot(j<m)$ only.
%

%
%The above example indicates that while the LLM can capture the main idea of the program, i.e., the loop invariants it proposed are almost correct. But it fails on some details so that the loop invariant cannot be verified formally.
This initial result highlights a key challenge: while the LLM grasps the program's main logic, it may fail on critical details required for a formal check.
Our goal is to provide the LLM with targeted feedback on such details that help it correct its answer.
%
%Here, a trivial approach would be simply telling the LLM that your previous proposed loop invariants are all correct, but are not sufficient to verify the assertion.
Here, a trivial feedback strategy would be simply stating that the verification condition is wrong, such as ``the invariants you proposed are correct but insufficient to prove the assertion".
However, this naive strategy is ineffective:
in our experiments, this led the LLM to substitute $-m \leq a\leq m$ for the loop invariant $i1$ as a new one.
%
% OLD: While this property is what we want to prove, it is too strong to be a valid inductive invariant, because its preservation condition cannot be proved.
% Is `strong` here accurate?
While this property is what we want to prove, it is not a valid invariant because its preservation condition cannot be proven.
Therefore, more precise feedback is needed - one that identifies the specific flaw in the LLM's thinking process.
\begin{figure}
    \centering
    \begin{minipage}{0.9\textwidth}
    \begin{lstlisting}
- From invariant **i1**: `a >= -(j - 1) && a <= (j - 1)`
  - At the end of the loop, `not B` implies `j > m`.
  - Substituting `j = m + 1`:
    - At termination, `a >= -(m + 1 - 1) && a <= (m + 1 - 1)`
    - Simplifies to: `a >= -m && a <= m`.
    \end{lstlisting}
    \caption{A local reasoning step of the natural language proof given by the LLM.}
    \label{fig:example-natural}
    \end{minipage}
    \vspace{0.2in}

    \begin{minipage}{0.9\textwidth}
    \begin{lstlisting}
[Initial]
a >= -(j - 1) && a <= (j - 1)
j > m

[Proof]
(j > m) ==> (j == m + 1) // At loop termination, j is m + 1.
(j == m + 1) && (a >= -(j - 1) && a <= (j - 1)) ==> (a >= -m && a <= m) // Substituting j = m + 1 into i1, the range for a becomes -m to m.

[Conclusion]
a >= -m && a <= m
    \end{lstlisting}
    \caption{The corresponding formalized proof of the local reasoning step.}
    \label{fig:example-formalized}
    \end{minipage}
    
\end{figure}

\paragraph{Getting Natural Language Proof.}
To pinpoint the error, we first prompt the LLM to generate a proof in natural language to demonstrate its thinking process.
Since the establishment and preservation of the loop invariants have been checked valid, we ask the LLM to focus on proving the post-condition.
That is, demonstrating that the proposed invariants imply the assertion after the loop terminates.
Figure~\ref{fig:example-natural} shows part of the proof given by the LLM.
In this proof, the LLM correctly deduces the loop exit condition is $j>m$.
However, it then incorrectly assumes $j=m+1$.
This step is a logical leap: without the upper bound on $j$, $j>m$ does not necessarily imply $j=m+1$.
\paragraph{Formalizing the Proof.}
%Next, we expect to detect the error occurs in the natural language proof. 
%
Next, to automatically and formally detect such reasoning errors, we formalize the LLM's natural language proof.
%In order to identify the proof formally and automatically, we want to formalize the natural language proof.
%
%A method to formalize the proof here is to let the LLM write a proof script in language such as Coq or Isabelle, but here comes some problems.
One may consider having the LLM generate a proof script in a language such as Coq~\cite{Coq97} or Isabelle~\cite{Isabelle02}.
%
%However, there are drawbacks to this approach.
%
%It shifts the task's focus to proof script generation, instead of detecting reasoning errors.
%
% OLD: \yannew{Due to the difficulty of generating a correct proof of these theorem provers,}
%\yannew{Due to the difficulty of correctly applying tactics of these theorem provers,}
%\todotcli{[REFINEMENT NEEDED]}
%
However, there is a gap between such formal language proofs and the natural language thinking process, and it is a difficult task for LLMs to generate a fully correct proof script.
The proof scripts generated by the LLM often fail due to non-logical reasons, such as wrong theorem application or incorrect rewrite~\cite{LuDZ24}.
This makes it difficult to isolate fundamental reasoning flaws from such errors.
%But what we want to do is to detect the errors in the LLM's reasoning process to guide the LLM toward correct solutions, instead of letting it write a complete formal proof which can be verified.
%
%Second, the proof scripts written by LLMs often fail due to the tactic issues (e.g., the misuse of tactics), making it hard to correspond the error in the proof script to that in the LLM's reasoninn process.
%

%
Instead, we use a more lightweight formalization technique.
The goal of our formalization is to detect the \textit{local} errors in the LLM's thinking process, which can then be used to guide it to ``guess'' a correct answer.
%In a new chatting session
Specifically, we provide an LLM with the natural language proof and instruct it to translate the key reasoning steps into a sequence of first-order logic implications.
As shown in Figure~\ref{fig:example-formalized}, the LLM identifies the initial conditions and expresses the reasoning process as two implications.
Each implication is annotated with a comment linking it back to the original proof step.
%
%A comment is attached to each implication so that the step in the original reasoning can be correlated.
%
%This gets rid of the LLM from generating complex formalized proof scripts.
This process avoids the complexities of generating complete proof scripts while producing formal claims that can be validated.
Using an SMT solver, we find that the first implication, $(j>m)\implies (j=m+1)$, {is} invalid.
This step successfully identifies the precise location of the logical flaw in the LLM's thinking process.
%
%The validity of an implication can be checked by an SMT solver.
%
%Besides, we can also check other things like whether the LLM uses correct conditions in a reasoning step.
%
%We will explain what we check in Section~\ref{sec:approach} in detail.
%
%In this example, the first implication can be checked as invalid by an SMT solver, because $j>m$ cannot derive $j=m+1$. 
%
%Therefore, the error is now identified.
%

%
\paragraph{Feedback to the LLM}
With the error identified, we construct precise feedback.
We provide a new prompt to the LLM, stating: ``In your reasoning, the step `At loop termination, $j$ is $m + 1$' is wrong. This is because the condition $(j > m)$ cannot derive $(j == m + 1)$''.
The feedback includes where the LLM makes mistakes and how it gets wrong.
We then instruct it to reconsider its loop invariants in light of the flaw of its reasoning.
Based on this feedback, the LLM produces the refined invariants in Figure~\ref{fig:example-refine}.
It correctly adds the missing upper bound for $j$ ($j\leq m+1$ as $i3$) without altering the other correct invariants.
With this complete set of invariants, the verification task succeeds.

%After identifying the error, we can provide it as feedback to the LLM.
%
%The feedback contains where the LLM made mistakes in its reasoning and how it made mistakes.
%
%In this example, we can get the location where the LLM made mistakes from the comment attaching to the implication.
%
%We prompt to the LLM, ``In your reasoning ,the step `At loop termination, j is m + 1' is wrong. This is because the condition (j > m) cannot derive (j == m + 1).''
%
%Additionally, we provide instructions to help the LLM refine its answer.
%
%We ask it to first carefully think why the error occurs here, and based on this, refine its proposed loop invariants.
%
%Figure~\ref{fig:example-refine} shows the refined invariants.
%
%This time, the LLM adds the upper bound of $j$ without changing the other ones.
%
%The loop invariants are now correct and the verification task is completed.

\section{Preliminaries}
\label{sec:pre}

The problem we focus on is loop invariant synthesis. Specifically, we focus on synthesizing loop invariants to verify the post-loop condition for programs with a single loop. Formally, consider a Hoare triple over a loop $\{P\}\ \texttt{while }B\texttt{ do }S\ \{Q\}$, which means that $P$ is the pre-condition that holds before the loop, and $Q$ is the post-condition that holds after the loop if the loop terminates. By the classical Hoare Logic~\cite{Hoare1969}, we have

\[
\frac{P\implies I\quad \{I\land B\}\ S\ \{I\}\quad (I\land\lnot B)\implies Q}{\{P\}\ \texttt{while }B\texttt{ do }S\ \{Q\}}
\]

This indicates that to prove the Hoare triple $\{P\}\ \texttt{while }B\texttt{ do }S\ \{Q\}$, we need to find an inductive loop invariant $I$, which satisfies the following three conditions:

\begin{enumerate}
    \item \textit{Establishment.} The loop's pre-condition $P$ implies the loop invariant $I$ $(P\implies I)$.
    \item \textit{Preservation.} Given the loop invariant $I$ holds before one iteration of the loop, and the loop executes for one more iteration, $I$ must also hold at the end of the loop iteration $(\{I\land B\}\ S\ \{I\})$.
    \item \textit{Post-condition}. If the loop exits, the loop invariant must imply the post-condition $Q$ $((I\land\lnot B)\implies Q)$.
\end{enumerate}

If the above three conditions are satisfied, we say the loop invariant $I$ is an \textit{inductive loop invariant}, and the program property $Q$ can then be proved using $I$.
\section{Overall Framework}
\label{sec:overall}

\begin{algorithm}[t]
\caption{Iterative Invariant Refinement Framework.}
\label{alg:framework}
\begin{algorithmic}[1]
\REQUIRE A single-loop program $P$, an assertion $a$, an invariant synthesizing LLM $L_S$, a formalizing LLM $L_F$, an invariant verification tool $V$.
\ENSURE A set of inductive loop invariants \textit{inv}.

\STATE \textit{inv} := getInitialInvariants($L_S$, $P$, $a$)
\WHILE{\TRUE}
    \STATE $invalidVC$ := checkInvariants($V$, $P$, $a$, \textit{inv})
    \IF{$invalidVC$ is None}
        \STATE \textbf{break}
    \ENDIF
    \STATE \textit{naturalProof} := getNaturalProof($L_S$, $P$, $a$, \textit{invalidVC})
    \STATE \textit{formalizedProof} := formalizeProof($L_F$, \textit{naturalProof})
    \STATE \textit{errors} := checkProof(\textit{formalizedProof})
    \STATE \textit{inv} := getRefinedInvariants($L_S$, $P$, $a$, \textit{inv}, \textit{errors})
\ENDWHILE
\RETURN \textit{inv}
\end{algorithmic}
\end{algorithm}

Algorithm~\ref{alg:framework} demonstrates the overall framework of our approach.
It proposes loop invariants to verify the assertion $a$ in a single-loop program $P$.
An LLM $L_S$ is used to synthesize the loop invariants, which we refer to as the \emph{Synthesizer LLM}. And an LLM $L_F$ is employed to formalize the natural language given by $L_S$, which we refer to as the \emph{Formalizer LLM}.
%
%\yannew{($LI$ and $LF$ do not need to be different.)}
%
The loop invariants and the assertion can be formally checked by the verification tool $V$.
The algorithm first prompts the Synthesizer LLM $L_S$ to propose invariants directly. The invariants are then checked by $V$.
If the invariants are checked to be inductive, the verification task is completed.
Otherwise, the verification tool will return a set of invalid verification conditions, and the algorithm goes into a feedback-refinement iteration.
In each iteration, the Synthesizer LLM is first asked to give a step-by-step proof for the verification condition that fails.
Then, the algorithm prompts the natural language proof to the Formalizer LLM $L_F$ to let it formalize the proof into a sequence of first-order logic implications.
The formalized proof can be checked for errors, which are then provided as targeted feedback to guide the Synthesizer LLM to correct the errors and propose refined loop invariants.
The iteration proceeds until the inductive loop invariants are found.

%\yannew{How about programs with more than 1 loop?}

%\todotcli{Our approach currently considers programs with one loop and one method. Maybe we should state this in the preliminary section?}

\section{Our Approach}

%\todotcli{TCLI: I'm working on refining this section. Please review other sections first before this comment is removed.}

In this section, we introduce the design of our framework in detail.
Specifically, we introduce our approach in the following steps: the primitive ``guess-and-check'' paradigm and the formal feedback and refinement mechanism. Our approach iteratively applies these two steps until a set of inductive loop invariants is generated.

\subsection{The Primitive Guess-and-Check Paradigm}
% We first apply a primitive ``guess-and-check'' paradigm to synthesize loop invariants.
%
In the primitive ``guess-and-check'' paradigm, we first ask the Synthesizer LLM to generate a set of candidate loop invariants. Subsequently, we leverage a formal verification tool $V$ to check whether the generated invariants are correct and sufficient to prove the desired program assertion.
%

\begin{figure}
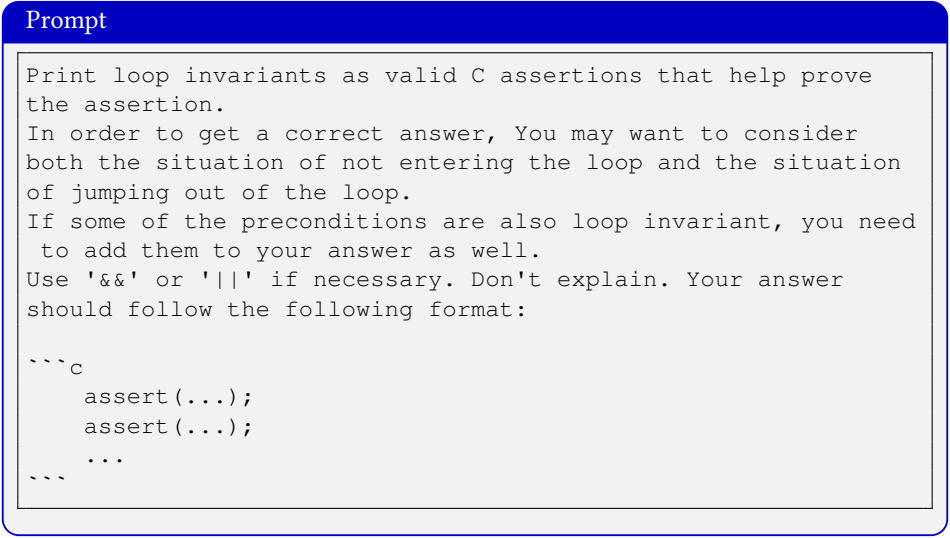

    \centering
    \begin{minipage}{0.9\textwidth}
    \begin{PromptBox}
Print loop invariants as valid C assertions that help prove the assertion. 
In order to get a correct answer, You may want to consider both the situation of not entering the loop and the situation of jumping out of the loop. 
If some of the preconditions are also loop invariant, you need to add them to your answer as well. 
Use '&&' or '||' if necessary. Don't explain. Your answer should follow the following format:

```c 
    assert(...);
    assert(...);
    ...
```
    \end{PromptBox}
    \caption{Prompt for the Synthesizer LLM to generate initial invariants.}
    \label{fig:LI-prompt}
    \end{minipage}
\end{figure}

%
%Specifically, at the first step, we prompt the LLM $LI$ to propose a set of loop invariants $\mathbb{L}$ given a program $P$ and an assertion $a$. Our prompts are shown in Figure~\ref{fig:LI-prompt}.
Specifically, the process begins by providing the Synthesizer LLM with the program source code $P$ and a post-condition assertion $a$. 
We use a carefully crafted prompt, shown in Figure~\ref{fig:LI-prompt}, to guide the LLM in proposing a set of loop invariants, $\mathbb{L}$.
The model is expected to generate invariants that are not only valid but also strong enough to imply the assertion $a$.

For the ``check'' step, we use Frama-C~\cite{CuoqKPSSY12} as the formal verification tool $V$ to perform the verification. By building on Frama-C, the generated invariants are ensured valid for a C program.
%to perform a two-step verification.
%
%First, we check whether the invariants hold before and after the loop.
%
%Second, we check whether the invariants imply the assertion $a$.
%
%Specifically, we rewrite the loop invariants given by the LLM into the annotations as shown in the example Figure~\ref{fig:example-propose}, which can be then checked by Frama-C.
%
Frama-C verifies the loop invariants and the program assertion by decomposing the overall goal into a set of verification conditions (VCs).
%
%If the verification passes, we can conclude that the invariants are correct and sufficient to prove the assertion.
%
Specifically, for a program $P$ with a single loop, a candidate invariant set $\mathbb{L}$ for the loop, and an assertion $a$ after the loop,  Frama-C generates VCs corresponding to the three standard proof obligations of Hoare Logic:
\begin{enumerate}
    \item \textbf{Establishment of Invariants:} For each loop invariant $l \in \mathbb{L}$, a VC is generated to verify that $l$ holds true before the loop, given the program's pre-conditions.
    \item \textbf{Preservation of Invariants:} For each loop invariant $l \in \mathbb{L}$, a VC is generated to verify that if all invariants in $\mathbb{L}$ hold at the beginning of an arbitrary loop iteration, $l$ will continue to hold after that iteration completes.
    \item \textbf{Implication of Assertion:} For the program assertion $a$, a VC is generated to verify that upon loop termination, the conjunction of all invariants in $\mathbb{L}$ is sufficient to imply the program assertion $a$.
\end{enumerate}

%Note that each of the above VCs is independent, i.e., for each verification condition checking, Frama-C assumes the conditions in the check hold, although they might not be proved in other VCs.
Note that Frama-C treats each verification condition as an independent proof obligation.
During the verification of a given VC, its premises are assumed to hold, regardless of the proof status of other VCs.
%
%This allows for a focused analysis of individual proof failures later on.
This modularity is crucial for pinpointing the exact sources of failure later in our feedback process.
For each VC, Frama-C returns a success or failure result.
%
%If all VCs pass, we conclude that the invariants are correct and sufficient to prove the assertion, thus the verification task is completed.
If all VCs pass the validation, the verification task is considered successful, which confirms that the loop invariants are both correct and sufficient to prove the assertion.
Otherwise, we will get a set of failed VCs (denoted as $\mathbb{VC}_{fail}$), which indicates that either some invariants are incorrect (i.e., they fail the establishment or preservation checks) or the current invariants are not sufficient to prove the final assertion.
% \yannew{or the preservation of those invariants}.
%Frama-C will return a pass/fail result for each VC. If all VCs pass, we can conclude that the invariants are correct and sufficient to prove the assertion. Otherwise, we will get a set of failed VCs (denoted as $\mathbb{VC}_{fail}$), which indicates that either some invariants are incorrect or current invariants are not sufficient to prove the assertion.

\subsection{Formal Feedback and Refinement}
\label{sec:approach-feedback}
While the primitive ``guess-and-check'' paradigm can succeed on simple problems, it often fails when the invariants are complex.
A key capability of modern LLMs is their ability to refine their outputs based on feedback~\cite{kamoi-etal-2024-llms}.
However, the nature of the feedback is critical.
%
%In addition to the primitive ``guess-and-check'' paradigm, as LLMs have the ability to correct their own mistakes by receiving feedback\morecite{}, we intend to provide feedback to the LLM for refinement.
%
The prior work~\cite{AdharshKSLDRR23} has explored using binary (success/failure) feedback to guide the LLM to refine its loop invariants.
This approach offers a weak signal, informing the LLM only \textit{that} its proposal is incorrect, but not \textit{why}.
It fails to target the root cause of the failure, which is often a subtle logical error or hallucination rather than a complete misunderstanding of the program.
As a result, the binary feedback leads to a limited improvement on the LLM.
%
%The previous work~\cite{AdharshKSLDRR23} also considers feedback, but their feedback is binary feedback with only success or failure, which leads to limited improvement for the LLM\yannew{, and does not target the root cause of those failure: hallucination or logical errors}.
%
%Besides, as shown in Section~\ref{sec:example}, LLMs can often capture the main behavior of the program, but fail on local, detailed logical errors.

%
To overcome this limitation, we propose a formal feedback and refinement mechanism to automatically diagnose the LLM's thinking process and provide targeted feedback.
This mechanism forms the core technical contribution of our work.

%

%However, as LLMs are not perfect, they usually perform poorly on complex tasks, such as intruct them to generate a proof that are both correct and syntextically valid for formal proof scripts such as Coq or Isabelle, we split the task into three steps:
%
Our formal feedback and refinement framework proceeds in the following steps:
\begin{enumerate}
    %\item \textbf{Structured Natural Language Proof Generation:} We ask the LLM to generate a step-by-step \yannew{thinking process} to prove the correctness VCs in $\mathbb{VC}_{fail}$ in structured natural language.
    \item \textbf{Structured Natural Language Proof Generation:} We prompt the Synthesizer LLM to generate its thinking process by generating a step-by-step natural language proof for a failed VC.
    \item \textbf{Formalization and Checking:} We translate the Synthesizer LLM's thinking process generated in the above step into a series of formal logical expressions using the Formalizer LLM, and use an SMT solver to automatically verify each reasoning step.
    \item \textbf{Feedback and Refinement:} We synthesize the results of the formal check into a report that pinpoints the exact logical errors, which is then provided to the Synthesizer LLM to guide the generation of a refined set of invariants.
\end{enumerate}

\subsubsection{Structured Natural Language Proof Generation}

% goals: effective, prepared for formalization
%In this step, we aim to make the LLM $LI$ generate a \textbf{Detailed} natural language proof to make it can be checked by a verifier, and the proof should also be easily \textbf{Formalizable} into a set of formal logical expressions.
%
%The goal of this step is to let the LLM $LI$ generate a \textbf{Detailed} natural language proof which can reflect its thinking process, and the proof should also be conveniently \textbf{Formalizable} into a set of formal logical expressions. 
The goal of this step is to obtain a natural language proof from the Synthesizer LLM that is both \textit{detailed} enough to expose its underlying thinking process and structured in a way that is \textit{ready to be formalized}.

%However, on the one hand, \liunew{existing works or our experiments?? IDK} show that LLMs can only generate a coarse-grained proof when the proof a goal consists of multiple \liunew{sub-goals(maybe a more precise expression)}~\morecite{}.
%
However, our initial experiments revealed that prompting the LLM to justify the entire verification task at once often results in a coarse-grained proof, which tends to merely re-states the high-level VCs.
%However, in our experiments, we found that the LLM tends to generate a coarse-grained proof if we let it to generate the whole thinking process of the overall verification task.
%
%For example, it only repeats the verification conditions to show its thinking, which cannot be checked for detailed errors.
%
To obtain more detailed reasoning, our method focuses the LLM on a single failed VC from $\mathbb{VC}_{fail}$ at a time.
%
%To this end, we only ask the LLM to focus on one specific VC in $\mathbb{VC}_{fail}$ and give the proof at a time.
%To this end, we only ask $LI$ to generate a proof for one VC in $\mathbb{VC}_{fail}$ at a time.
The specific VC is selected following a natural deductive order: \emph{establishment of invariants, preservation of invariants, and then implication of assertion}. This order is the same as the standard methodology of proof in Hoare Logic.
If there are multiple invalid VCs of the same type, a random one is picked.
%
%Specifically, we select VC following the order of \emph{establishment of invariants, preservation of invariants, and implication of assertion}, and when there are multiple VCs of the same type, we randomly select one VC. 
%This is the same order as a natural proof process such as which is used in Hoare Logic~\cite{Hoare1969}.
%

\begin{figure}
    \centering
    
    \includegraphics[width=0.9\textwidth]{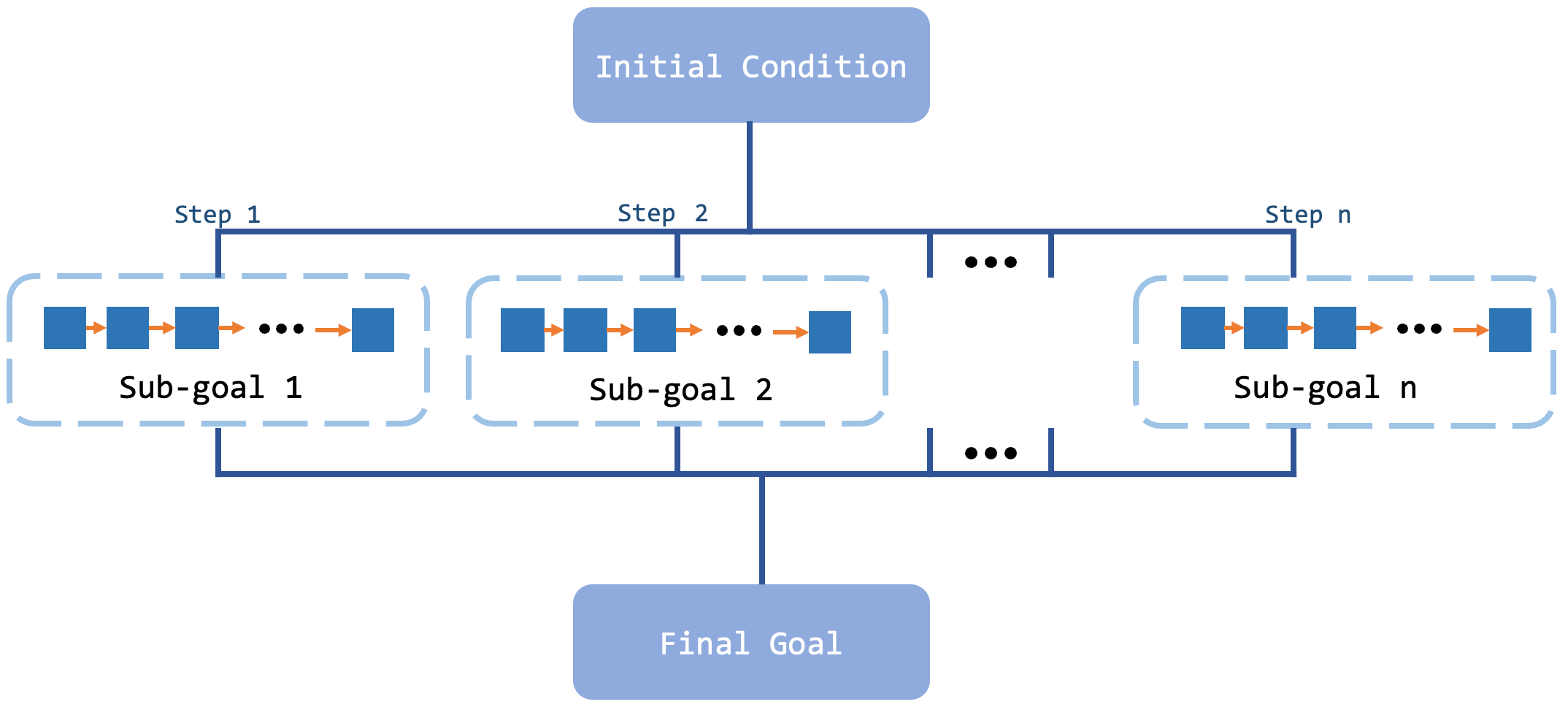}
    \caption{The structure of the natural language proof.}
    \Description{A diagram showing the structure and steps of a natural language proof as generated by the LLM.}
    \label{fig:proof-structure}
\end{figure}

%
To ensure the proof is ready to be formalized, we constrain its structure using the prompt shown in Figure~\ref{fig:proof-prompt}.
The LLM is instructed to first list all initial conditions from the VC, and then decompose its reasoning into a sequence of numbered steps, as depicted in Figure~\ref{fig:proof-structure}.
Each step is expected to represent a single, sequential line of reasoning.
In other words, if a sub-goal requires case analysis, the LLM is instructed to split the cases into separate steps.
This structured format allows each step's core logic to be easily extracted and formalized.
%
%On the other hand, to make the proof easily formalizable, we prompt the natural language proof to be well structured.
%
%Specifically, we prompt the LLM $LI$ to generate proofs following the step-by-step format shown in Figure~\ref{fig:proof-structure}. \todotcli{Figure need to be repaired.}
%
%We ask $LI$ to first list the initial conditions, which can be directly obtained from the VC. 
%Then $LI$ can only perform a one-layer classification discussion. In each case, $LI$ can only use a sequential reasoning process. 
%Then, it is asked to decompose the final goal into a sequence of steps which are explicitly marked with numbers.
%
%In each step, the LLM is expected to focus on a sub-goal which can be proved in a sequential reasoning.
%
%In other words, if a sub-goal has to be proved by cases, the LLM should split the cases in different steps.
%
%Consequently, the proof of each step can be easily formalized by our following formalizing technique.
%
As an example, Figure~\ref{fig:example-natural} presents a single step of the natural language proof given by the LLM, which performs a sequential reasoning from the known conditions to the program assertion.
In the checking process, our method formalizes and verifies the reasoning within each step.
It does not formally verify the high-level case-split structure of the proof itself.
The primary goal of our work is to detect and correct the local, detailed logical errors in the LLM's thinking process.
As our experiments in Section~\ref{sec:evaluation} show, this targeted focus on local reasoning errors is highly effective at guiding the LLM toward a correct solution.
%However, the purpose of our work is to find local and detailed reasoning errors of the LLM to guide it to a correct solution.
%
%In our experiment, we show that such a checking process is sufficiently effective to guide the LLM for refinement.
%However, we only formalize and check the proof inside each case, and we do not check the classification discussion itself, so it may lose some procise. However, in practice, we find that this is sufficient to guide the LLM to generate correct invariants.
%

%Our prompt for generating the structured natural language proof is shown in Figure~\ref{fig:proof-prompt}.

\begin{figure}
    \centering
    % \begin{minipage}{0.9\textwidth}
    \begin{PromptBox}
 Here are some instructions on step-by-step Structure for your proof:
 - The whole proof contains three parts: initial conditions, proof process, and conclusion. Label them with `[Initial]`, `[Proof]`, and `[Conclusion]`, respectively.
 - In the `[Initial]` part, just list the initial conditions, do not perform any logical inference - leave them to the proof process.
 - In the `[Proof]` part, label each step explicitly with the format:
  `[STEP N: <STEP_NAME>]` where `N` is the step number and `<STEP_NAME>` describes the step. Do not miss the left `[` and the right `]`.
 - Under each step, provide detailed mathematical/logical reasoning.
 - Each step should be a sequential logical reasoning. For example, if a step contains subcases, please split them into different steps.
    \end{PromptBox}
    % \end{minipage}
    \caption{Prompt for the Synthesizer LLM to generate structured natural language proof.}
    \label{fig:proof-prompt}
\end{figure}

%
% Introduce another .tex file to avoid conflicts.
\subsubsection{Formalization and Checking}

%{\paragraph{REMINDER} For things I am not sure, I use \yannew{this format} to highlight them. Please pay more attention when reviewing them.}

%
%Given those proof steps provided by the $LI$, the next procedure is to prove or disprove them so that we could generate the feedback to the $LI$.
%
%G
Given the structured natural proof from the Synthesizer LLM, the next task is to formally check it for logical errors.
To maintain a clean context separation between loop invariant generation and formalization, we utilize a separate LLM session, the Formalizer LLM, for this translation task.

%Given the natural language proof steps given by the loop invariant generation LLM, the next procedure is to check the its errors formally so that we can provide targeted feedback.
%
%Since the context lengths of LLMs are limited and we do not want to mix contexts of invariant generation and formalization,
%we use a new session $LF$ to separate these 2 contexts of tasks.
%

%
Previous auto-formalization works~\cite{jiang2023draft,zhou2024dont} focused on targetting the LLM to generate a proof script in languages such as Coq or Isabelle, which can be checked for validity formally and completely.
However, LLMs often struggle to generate code that is correct in both logic and syntax.
It has been studied that the Coq code generated by an LLM often fails due to low-level issues  like wrong theorem application or syntactic mistakes~\cite{LuDZ24}.
Besides, the failures in the proof script often mask the actual error in the natural language thinking process.
For example, if the Coq code tries to apply the hypothesis \texttt{H: m=n} to \texttt{n=m}, it will fail with the error message ``Unable to unify `m=n' with `n=m'.'', which does not provide any information for the error in the thinking process.
Therefore, we cannot construct effective feedback based on the error of the proof script.
To address these issues, we take a  more lightweight formalization strategy.
Note that the goal of our approach is not for the LLM to write a completely verifiable proof script --- the final ``check'' is already handled by Frama-C.
Instead, we only need to detect local errors in the LLM's thinking process to guide its next ``guess''.
%We only need to detect the local errors of LLM's thinking process.
%
As a result, we prompt the Formalizer LLM to transform the key inference within each proof step into a series of first-order logic implications. 
%into a series of first-order logic implications which can be checked by an SMT solver.
%
This approach has two main advantages.
%We apply this method mainly for two reasons.
%
First, in practice, LLMs are significantly more successful at generating syntactically correct expressions in this simple format.
Second and more importantly, each implication maps directly back to a specific reasoning step in the natural language proof, which enables precise error attribution.
%
%We can then point out where the LLM's thinking process makes mistakes based on the invalid implication.
%

%
The formalization consists of three key parts:

\begin{itemize}
    \item A list of  all conditions assumed to be true at the beginning of the proof step.
    \item A sequence of first-order logic implications that represent the core reasoning. Each implication is annotated with a comment linking it to the original natural language step (see Figure~\ref{fig:example-formalized}).
    \item  The final conclusions derived in the step.
\end{itemize}

To facilitate automated checking, all conditions and conclusions  should be valid C logical expressions.
Since the checking can be conducted automatically and rigidly by the checker, we ask the LLM to focus on 
\emph{translating} instead of checking possible problems in those implications.
The prompts used for the formalization task are shown in Figure~\ref{fig:prompt-formalization1} and Figure~\ref{fig:prompt-formalization2}.

% Keep invariants alone for a clearer architecture!
\begin{figure}
    \begin{PromptBox}
Please read the above proof for one certain goal. For each step marked with [STEP N: ...] in the above proof, formalize its logical content into first-order logic expressions. Follow these rules:

1. When formalizing a certain step, focus on it. Formalize **only** the certain step, and ignore other steps.
2. List the known conditions before this step. Each condition should be a logical expression represented in C grammar. The detailed known conditions are listed as follows:
    - The initial conditions.
    - The conclusions that derived from previous big steps.
    - If this step needs declare a new variable, list it here using "==" symbol.
3. Formalization requirements:
    - Convert natural language claims into precise first-order logic formulas.
    - For each formula, represent it as an implication, using "==>" to represent.
    - The condition and the conclusion of the implication should be legal C logical expressions, i.e. it can only contain "&&", "||" and "!" as logic operators.
    - Each formula should be self-contained so that it can be checked.
    - You do not need to verify the correctness of the proof, just formalize it in logic formulas.
    - At last, summarize the conclusion of this step using a C logical expression. **Here, the conclusion means the one that are actually derived from this step, not the one that need to be derived.**

---

    \end{PromptBox}
    \captionof{figure}{The Prompt Used for Formalization (Part 1 / 2)}
    \label{fig:prompt-formalization1}
\end{figure}
\begin{figure}
    \begin{PromptBox}
Output format:
    - For each step marked with [STEP N: ...], first output the step label in one line. It should be the same as the label in the natural proof. Then, output the formalization of this step in the following format.
    - First, list the known conditions. This part begins with a `[Initial]` label, with one condition per line. Each condition should be a legal C logical expression. Each condition follows by a comment starting with "//". Follow the following rules for comments:
        1. If the condition comes from initial conditions, comment "initial".
        2. If the condition comes from conclusions derived by previous big steps, comment "derived".
        3. If the condition is a declaration of a new variable, comment "declaration".
        4. Comment a single word for each condition.
    - Then, output the formalized proof. This part begins with a `[Proof]` label, with one implication per line.
    - Output one implication in each line, following by a comment starting with "//" which matches the claim in the natural language proof. The condition and conclusion must be legal C logical expressions. For example: 
      ```
      (a <= 0) && (a == m) ==> (m <= 0) // From a is non-positive, and a is equal to m, we know that m is non-positive, too.
      ```
    - At last, output the conclusion of this step. This part begins with a `[Conclusion]` label. The conclusion stays within one line, with exactly one C logical expression. Similar to the above, represent the explanation to the conclusion in a comment starting with "//".
    - Only output the conditions, implications, conclusions,  and comments, do not explain.
    - For all C logical expressions mentioned above, do not use functions in it.

    \end{PromptBox}
    \captionof{figure}{The Prompt Used for Formalization (Part 2 / 2)}
    \label{fig:prompt-formalization2}
\end{figure} % Make the tex file clean!

%Given those formal logical expressions generated by session $LF$,
%the next procedure is to prove or disprove their correctness.

Figure~\ref{fig:example-formalized} presents an example of this kind of formalization.
In the example, two reasoning steps are transformed into two distinct logical implications.
Each implication is mapped with the original natural language proof step using a comment.
The formalized proof is then checked automatically.
If the focused verification condition is an establishment VC, we first check whether the initial conditions given by the LLM hold by integrating them as assertions in the original program and checking them by Frama-C.
This is because sometimes the LLM may incorrectly identify the pre-conditions, such as wrongly assume the range of an undefined variable.
Next, we check the sequence of implications using the procedure in Algorithm~\ref{alg:implication-checking}. 
Specifically, we maintain a set of known fact $Conds$ which is first initialized with the pre-conditions of the proof step given by the LLM.
Then, we scan over the implications in the list of implications $Imp$.
For each implication $p\Rightarrow q$, we first check if its premise $p$ is entailed by the current set of known facts ($Conds \models p$).
%
%If not, we get that $p$ is an unsupported assumption, and add it to 
If not, it suggests the LLM uses an incorrect assumption in this specific step.
Then, we check the validity of the implication $p\Rightarrow q$ itself using an SMT solver.
After checking an implication, we add its conclusion $q$ to $Conds$ for subsequent steps, regardless of whether the implication is valid. In other words, if a step applies a wrongly-derived conclusion of the previous steps, we do not treat this as an error.
This allows our system to identify multiple, independent errors within a single proof attempt.
%

%
%After the checking, we add $q$ to the known condition set $Conds$, regardless of whether this implication holds.
%

%For the establishment part, 
%we simply insert its initial conditions into the original program and then use Frama-C to check their correctness.
%For the preservation part, there is a chicken-and-the-egg problem:
%Sometimes LLM proposes a correct invariant $I$, but $I$ maybe too weak to derive itself ($I\not\imply I$).
%To takcle this problem, we check those implications in an iterative way as shown in Algorithm~\ref{alg:implication-checking}.
%\yannew{The reason we add $q$ into $Conds$ is to minimize the set of invalid equations.
%Because sometimes LLM proposes valid antecedents but the proof is wrong.}

\begin{algorithm}[t]
\caption{Iterative Checking of Implications.}
\label{alg:implication-checking}
\begin{algorithmic}[1]
\REQUIRE A set of pre-conditions $Pre$, a list of implications $Imp$.
\ENSURE A set of invalid expressions $Invalid$.

\STATE $Conds$ := $Pre$
\STATE $Invalid$ := $\emptyset$
\FOR{ $p\imply q \in Imp$}
    \IF{$Conds\not\models p$}
        \STATE $Invalid$ := $Invalid\cup\{p\}$
    \ENDIF
    \IF{$p \not\models q$}
        \STATE $Invalid$ := $Invalid\cup\{p\imply q\}$
    \ENDIF
    \STATE $Conds$ := $Conds\cup\{q\}$
\ENDFOR
\RETURN $Invalid$
\end{algorithmic}
\end{algorithm}

\subsubsection{Feedback}

The final step synthesizes the set of invalid expressions, $Invalid$, into a targeted,  diagnostic feedback prompt for the Synthesizer LLM.
The goal is to clearly explain the errors to guide the generation of a corrected set of invariants.

The feedback is structured to be maximally informative. For each error found, we provide the LLM with:
\begin{enumerate}
    \item The context: The specific verification condition (e.g., preservation of invariant `i <= n') that was being analyzed.
    \item The location of the error: The specific step number from its natural language proof (e.g., `[STEP 3: ...]') and the reasoning location (given by the comment of the invalid implication).
    \item The specific logical flaw: The formal implication that was found to be invalid, or the initial conditions which are not satisfied.
    \item The nature of the flaw: A clear, templated explanation of whether the error was an unsupported assumption (the premise was not provable from prior facts) or a logical non-sequitur (the implication itself does not hold).
\end{enumerate}

An example of this feedback prompt is shown in Figure~\ref{fig:prompt-feedback}. By pinpointing the exact location and nature of the error in the thinking process, we provide a constructive feedback, moving from simple binary feedback to actively guide the LLM's refinement process.

%Given a set of local errors generated by the above step, we use them to form a feedback to the LLM for a refined version of invariants.
%
%To better guide the LLM to the issues it made, we tell it the step that contains the issues, the implication, and how it made those issues (whether use an invalid antecedent or the implication cannot hold).
%
%An example prompt is shown in Figure~\ref{fig:prompt-feedback}.

% Keep invariants alone for a clearer architecture!
\begin{figure}
    \centering
    
    \begin{PromptBox}
In your proof of {step}, "{comment}" is wrong. This is because the condition {condition} cannot derive {conclusion}.

This indicates that the loop invariants you proposed are all correct, but they are not sufficient to verify the assertion {assertion}. 

Please fix your loop invariants in the following steps:
First, based on the error in your proof, analyze why the loop invariant are not sufficient to verify the assertion {assertion}.
Then, from your analysis, carefully think about the verification problem again. Every time you propose a loop invariant, note that do not make the same mistake again.

In the end, print loop invariants as valid C assertions that help prove the assertion. 
In order to get a correct answer, You may want to consider both the situation of not entering the loop and the situation of jumping out of the loop. 
If some of the preconditions are also loop invariant, you need to add them to your answer as well. 
Use '&&' or '||' if necessary. Your answer should follow the following format:

```c 
    assert(...);
    assert(...);
    ...
```
    \end{PromptBox}
    \caption{The Prompt Used for Feedback.}
    \label{fig:prompt-feedback}
\end{figure} % Make the tex file clean!
\section{Evaluation}
\label{sec:evaluation}

%\todotcli{This section is currently written roughly. Please write or review other sections first before this comment is removed.}

In this section, we evaluate the effectiveness of our approach on a main benchmark suite of 460 C programs and a non-linear benchmark suite of 50 programs.
We aim to answer four research questions concerning the effectiveness, efficiency, and mechanisms of our proposed method.

\subsection{Evaluation Setup}

\paragraph{Baselines}

We compare \toolname{} with \baselinename{}~\cite{LaM4Inv} and \clausetoinv{}~\cite{Clause2Inv}, both of which represent state-of-the-art LLM-based approaches for loop invariant synthesis.
Both baseline methods operate by prompting an LLM to generate candidate invariants (or invariant components) and subsequently employ symbolic methods to get a correct set of invariants from the LLM-generated answer.
Specifically, \baselinename{} iteratively prompts an LLM for loop invariants, then validates these candidates and provides feedback to the LLM in the form of counterexamples generated by an SMT solver.
Its key technical contribution is the use of Bounded Model Checking (BMC) to filter and combine valid conjuncts from multiple, partially correct LLM responses.
\clausetoinv{} adopts a different strategy. Rather than generating full invariants, it prompts the LLM to produce a list of atomic clauses, which are then logically combined via conjunction or disjunction.
Like \baselinename{}, it relies on counterexample-based feedback to guide the LLM to generate new clauses.
Our approach diverges from these baselines in two fundamental ways.
First, while both baselines rely on coarse-grained, counterexample-based feedback, our approach provides structured and detailed feedback from formally verifying the LLM's step-by-step thinking process.
Second, while the baselines depend heavily on external symbolic methods to assemble the final solution from imperfect outputs, our approach focuses on enhancing the LLM's own reasoning capability to directly synthesize correct invariants.
Furthermore, our feedback mechanism is complementary to these symbolic techniques. We evaluate a potential combination between BMC used in \baselinename{} and our approach in RQ4.

\paragraph{Benchmark}

Our main benchmark suite consists of 460 C programs that require numerical invariants for verification, encompassing both linear and non-linear properties.
This suite is an aggregation of problems from several established sources to ensure diversity and difficulty.
The core of our benchmark consists of 313 programs evaluated for both \baselinename{}  and \clausetoinv{}  containing only linear properties.
We excluded 3 of the original 316 programs due to their use of floating point arithmetic, which is unsupported by the version of Frama-C we use.
To challenge our approach beyond the scope of the baselines, which have already achieved a near-perfect solve rate (309/316 for \baselinename{}, and 312/316 for \clausetoinv{}) on their own benchmarks, we incorporated more difficult problems.
We collected additional programs from the Microsoft loop invariant generation experiments~\cite{AdharshKSLDRR23}.
This dataset includes all the benchmarks from previous works - Code2Inv~\cite{Code2Inv18, Code2Inv20}, Accelerating Invariant Generation~\cite{MadhukarWKLS15}, and LinearArbitrary-SeaHorn~\cite{ZhuMJ18}.
Furthermore, we integrated benchmarks from all the sub-directories named with the prefix ``loop'' in the SV-COMP repository~\cite{svcomp}.
From these sources, we filtered for all programs containing a single loop and a single function, and removed duplicates to form our final suite of 460 programs.
In addition to the main benchmark suite, we also evaluate \toolname{} on the non-linear benchmark suite from the \clausetoinv{} paper.
This benchmark contains 50 more C programs each of which involves non-linear properties.

\paragraph{Evaluation metrics}

To account for the stochastic nature of LLMs, we execute each tool on every benchmark for five independent runs.
We consider the following evaluation metrics:

\begin{enumerate}
    \item \textbf{\# Solved Benchmarks:} The total number of unique benchmarks solved in at least one of the five runs.
    \item \textbf{Success Rate:} The average success rate, calculated as the total number of successful runs divided by the total number of attempts (i.e., $460\times 5=2300$ for the main benchmark suite).
    \item \textbf{Cost on Success:} The average number of tokens (input/output) and time (in seconds) consumed for successful runs only.
\end{enumerate}

To specifically analyze our contribution, we also distinguish between successful runs where the LLM provided a correct invariant in the first attempt (``direct solve'') versus those that required one or more rounds of our feedback mechanism (``feedback-driven solve''). We will explain them in detail in Section~\ref{sec:RQ3}.

%To avoid the effect caused by the random behaviors of LLMs, we run each problem five times independently.
%
%We consider the following evaluation metrics: 1) The number of solved benchmarks. If a benchmark is solved at least once in the five tries, then it is counted as a successful benchmark. 2) The average success rate. This metric is calculated by the successful tries divided by the total tries in the five evaluations. 3) The average token and time cost on the successful tries.
%
%For our approach, we also evaluate the successful tries that succeeded using our feedback method and the ones that succeed in one iteration. This is because in some tries, the LLM returns the correct answer in a simple one prompt, which cannot show the effectiveness of our method.

%
\paragraph{Implementation Details}

All of our experiments were conducted on a Linux machine with 256GB memory and 2.6GHz processors.
For fairness, we set a uniform time limit of 600 seconds and a token limit of 150,000 tokens per benchmark run for both our approach and the baselines.
We used Frama-C version 27.1 to verify the inductiveness of the loop invariants, with a timeout value of 5 seconds per verification condition.
%
%We set the timeout value of Frama-C to five seconds, a verification condition will be returned as ``unknown'' if the time exceeds.
%
%For the implications in the formalized proof, we use Z3 solver to check their validation.
The validity of logical implications in our formalized proofs was checked by the Z3 SMT solver~\cite{z3}.
We evaluated our approach on OpenAI's language models. We selected the most recent generative models with different parameter sizes, \texttt{gpt-4.1} and \texttt{gpt-4.1-mini}, and the older \texttt{gpt-4o-mini}.
We also considered one of the most recent and economical reasoning models \texttt{gpt-o4-mini}.
To demonstrate the effectiveness of our approach on non-OpenAI models, we further considered Anthropic's \texttt{claude-3.7-sonnet}.
For our approach, the same model was used as both the Synthesizer LLM and the Formalizer LLM.

%\todotcli{The LLMs evaluated include OpenAI's generation models \texttt{gpt-4.1} and \texttt{gpt-4.1-mini}, the older \texttt{gpt-4o-mini}, the reasoning model \texttt{gpt-o4-mini}, and Anthropic's \texttt{claude-3.7-sonnet}. For our approach, the same model is used for both invariant generation and proof formalization.}
%

%

\subsection{Evaluation Results}

We intend to answer the following research questions (RQs) with our experiments:

\begin{enumerate}
    \item \textit{Effectiveness}: How does \toolname{} compare to the baselines in terms of the number of solved benchmarks and overall success rate?
    \item \textit{Efficiency}: What is the time and token cost of our approach on successful runs?
    \item \textit{Impact of Feedback}: What is the direct contribution of our formal-verification feedback mechanism to the overall success?
    \item \textit{Combining with symbolic methods}: How does combining our approach with symbolic methods impact the verification performance?
\end{enumerate}

\subsubsection{RQ1: Effectiveness}

%\begin{table}[t]
%\caption{Overall effectiveness and cost comparison between \toolname{} and \baselinename{} across various LLMs. Costs are averaged over successful runs. Best results in each category are in bold.}
%\label{tab:effectiveness}
%\centering
%\resizebox{\textwidth}{!}{%
%\begin{tabular}{l|cccc||cccc}
%\toprule
%\multirow{2}{*}{\textbf{Model}} & \multicolumn{4}{c||}{\textbf{\toolname{}}} & \multicolumn{4}{c}{\textbf{\baselinename{}}} \\
%\cmidrule{2-9}
%& \textbf{\# Solved} & \textbf{Success Rate (\%)} & \textbf{Tokens (I/O)} & \textbf{Time (s)} & \textbf{\# Solved} & \textbf{Success Rate (\%)} & \textbf{Tokens (I/O)} & \textbf{Time (s)} \\
%\midrule
%GPT-4.1             & \textbf{445} & \textbf{93.1} & 8296 / 2738  & 55.9 & 414 & 84.6 & 4477 / 666   & \textbf{40.5} \\
%GPT-4.1-mini        & \textbf{439} & \textbf{91.1} & 10203 / 3923 & 52.8 & 412 & 84.2 & 3808 / 528   & \textbf{52.0} \\
%GPT-4o-mini         & \textbf{431} & \textbf{81.9} & 14302 / 5161 & 104.3& 397 & 76.0 & 6572 / 857   & \textbf{47.0} \\
%GPT-o4-mini         & \textbf{435} & \textbf{89.2} & 1557 / 4056  & \textbf{92.1} & 387 & 74.3 & 1890 / 12461 & 227.9 \\
%Claude-3.7-Sonnet   & \textbf{443} & \textbf{89.9} & 7155 / 2186  & \textbf{86.7} & 413 & 81.4 & 2425 / 1363  & 92.7 \\
%\bottomrule
%\end{tabular}
%}
%\end{table}

\begin{table}[t]
\caption{Effectiveness and cost comparison on the main benchmark suite (460 programs). \toolname{} is compared against \baselinename{} and \clausetoinv{}. Costs are averaged over successful runs. Best results for each model are in bold.}
\label{tab:effectiveness}
\centering
\resizebox{\textwidth}{!}{%
\begin{tabular}{c|lcccc}
\toprule
\textbf{Model} & \textbf{Method} & \textbf{\# Solved} & \textbf{Success Rate (\%)} & \textbf{Time (s)} & \textbf{Tokens (I/O)} \\
\midrule
\multirow{3}{*}{GPT-4.1}
 & \toolname{}    & \textbf{445} & \textbf{93.1} & 55.9 & 8296 / 2738 \\
 & \baselinename{}& 414          & 84.6          & 40.5 & 4477 / 666 \\
 & \clausetoinv{} & 409          & 84.9          & \textbf{12.4} & \textbf{2344} / \textbf{406} \\
\midrule
\multirow{3}{*}{GPT-4.1-mini}
 & \toolname{}    & \textbf{439} & \textbf{91.1} & 52.8 & 10203 / 3923 \\
 & \baselinename{}& 412          & 84.2          & 52.0 & 3808 / \textbf{528} \\
 & \clausetoinv{} & 405          & 83.6          & \textbf{18.4} & \textbf{3261} / 706 \\
\midrule
\multirow{3}{*}{GPT-4o-mini}
 & \toolname{}    & \textbf{431} & 81.9          & 104.3& 14302 / 5161 \\
 & \baselinename{}& 397          & 76.0          & 47.0 & 6572 / 857 \\
 & \clausetoinv{} & 408          & \textbf{82.6} & \textbf{20.8} & \textbf{3283} / \textbf{632} \\
\midrule
\multirow{3}{*}{GPT-o4-mini}
 & \toolname{}    & \textbf{435} & \textbf{89.2} & 92.1 & 1557 / \textbf{4056} \\
 & \baselinename{}& 387          & 74.3          & 227.9& 1890 / 12461 \\
 & \clausetoinv{} & 414          & 85.0          & \textbf{45.8} & \textbf{655} / 5026 \\
\midrule
\multirow{3}{*}{Claude-3.7-Sonnet}
 & \toolname{}    & \textbf{443} & \textbf{89.9} & 86.7 & 7155 / 2186 \\
 & \baselinename{}& 413          & 81.4          & 92.7 & \textbf{2425} / 1363 \\
 & \clausetoinv{} & 409          & 83.8          & \textbf{25.0} & 3645 / \textbf{812} \\
\bottomrule
\end{tabular}
}
\end{table}

\begin{table}[t!]
\caption{Effectiveness and cost comparison on the non-linear benchmark suite (50 Programs). \toolname{} is compared against \baselinename{} and \clausetoinv{}. Costs are averaged over successful runs. Best results for each model are in bold.}
\label{tab:effectiveness_nonlinear}
\centering
\resizebox{\textwidth}{!}{%
\begin{tabular}{c|lcccc}
\toprule
\textbf{Model} & \textbf{Method} & \textbf{\# Solved} & \textbf{Success Rate (\%)} & \textbf{Time (s)} & \textbf{Tokens (I/O)} \\
\midrule
\multirow{3}{*}{GPT-4.1}
 & \toolname{}    & \textbf{45} & \textbf{83.2} & 80.4 & 17365 / 5430 \\
 & \baselinename{}& 25          & 38.8          & 64.5 & 8334 / 1310 \\
 & \clausetoinv{} & 42          & 76.0          & \textbf{35.1} & \textbf{3915} / \textbf{741} \\
\midrule
\multirow{3}{*}{GPT-4.1-mini}
 & \toolname{}    & \textbf{46} & \textbf{80.8} & 74.1 & 11878 / 4573 \\
 & \baselinename{}& 29          & 44.8          & 73.2 & 8591 / 1383 \\
 & \clausetoinv{} & 42          & 73.2          & \textbf{24.3} & \textbf{2691} / \textbf{638} \\
\midrule
\multirow{3}{*}{GPT-4o-mini}
 & \toolname{}    & \textbf{44} & \textbf{77.6} & 125.2 & 12007 / 4407 \\
 & \baselinename{}& 27          & 36.0          & 60.9 & 8907 / 1188 \\
 & \clausetoinv{} & 42          & 70.4          & \textbf{40.3} & \textbf{4169} / \textbf{934} \\
\midrule
\multirow{3}{*}{GPT-o4-mini}
 & \toolname{}    & \textbf{47} & \textbf{87.6} & \textbf{61.8} & 1548 / \textbf{4802} \\
 & \baselinename{}& 30          & 48.8          & 222.8 & 3241 / 20218 \\
 & \clausetoinv{} & 46          & 83.6          & 61.9 & \textbf{785} / 5515 \\
\midrule
\multirow{3}{*}{Claude-3.7-Sonnet}
 & \toolname{}    & \textbf{47} & \textbf{79.2} & 88.0 & 13784 / 4752 \\
 & \baselinename{}& 29          & 44.4          & 95.7 & 11524 / 1965 \\
 & \clausetoinv{} & 43          & 78.8          & \textbf{37.9} & \textbf{4417} / \textbf{1125} \\
\bottomrule
\end{tabular}
}
\end{table}

Table~\ref{tab:effectiveness} and Table~\ref{tab:effectiveness_nonlinear} present the effectiveness of our approach compared to the baseline methods \baselinename{} and \clausetoinv{} on the main benchmark suite and the non-linear benchmark suite separately.
The results show that \textbf{our approach consistently outperforms the baselines in the number of solved benchmarks  across all the tested LLMs.}
On the main benchmark suite, using the most capable model, GPT-4.1, our approach solves 445 of the total 460 programs.
This is 31 more than \baselinename{} and 36 more than \clausetoinv{}, suggesting the effectiveness of our approach.
On the non-linear benchmark suite, our approach can solve at most 47 of the 50 programs using Claude-3.7-Sonnet.

Furthermore, \toolname{} achieves the highest overall success rate on four of the five tested models on the main benchmark suite.
%
%Besides, the overall success rate achieves 93.1\%, while that of baseline is only 84.6\%.
With GPT-4.1, our success rate reaches 93.1\%, significantly surpassing both \baselinename's{} (84.6\%) and \clausetoinv's{} (84.9\%).
This advantage holds for other state-of-the-art models like Claude-3.7-Sonnet and the more economical GPT-4.1-mini.
An exception is that with the weaker model, GPT-4o-mini, \clausetoinv{} achieves a slightly higher success rate (82.6\%) than \toolname{} (81.9\%).
This is because \clausetoinv{} prompts the LLM to generate simple clauses, a task that is easier for weaker models to perform consistently, leading to a higher success rate on easier problems.
However, our approach solves more unique benchmarks than \clausetoinv{} (431 vs. 408). This result shows that simple clauses fail to capture the complex invariants required for harder problems.
In contrast, our feedback mechanism effectively enhances the model's capability, allowing it to eventually solve a larger number of unique, difficult benchmarks that are beyond the reach of the baselines.
On the non-linear benchmark, \toolname{} achieves the highest success rate across all the tested LLMs, suggesting that our approach works better than the baselines on more difficult problems.
These outcomes support that providing detailed constructive feedback on the LLM's thinking process is more effective than providing only  a counterexample.
By pinpointing the exact logical flaws in a proposed proof, our approach enables powerful models to correct their thinking process and converge on a correct solution more reliably.
An interesting observation is that the advantage appears to scale with model capability.
Comparing the results of the three generative models of GPT series in Table~\ref{tab:effectiveness}, the performance gap between \toolname{} and the baselines on the success rate goes larger as the model becomes powerful.
%The performance gap between \toolname{} and \baselinename{} is the largest with the most powerful model GPT-4.1 (+31 solved).
%
%The trend also holds for the average success rate.
%
This suggests that as LLMs become more powerful, our approach is better equipped to leverage their enhanced abilities, hinting great potential for more advanced models in the future.

The results with the reasoning-specialized model, GPT-o4-mini, are particularly noteworthy when comparing with \baselinename{} on the main benchmark suite. While its absolute success rate (89.2\%) is lower than that of GPT-4.1, the performance gap between \toolname{} and \baselinename{} is the largest, at 14.9 percentage points. The lower absolute performance is partly due to the model's higher latency and token usage per turn, which limits the number of possible iterations within the fixed resource limits. However, the significant relative improvement demonstrates that \toolname{}'s structured feedback is highly compatible with the model's intended reasoning capabilities. On benchmarks solved via feedback, GPT-o4-mini requires an average of only 1.19 feedback rounds to find the solution, indicating that the feedback is precise and actionable.

\subsubsection{RQ2: Efficiency}

As shown in Table~\ref{tab:effectiveness} and Table~\ref{tab:effectiveness_nonlinear}, the efficiency trade-offs between the approaches are clear.
%
%On the generating models, our approach takes more tokens than the baseline, and the time cost is comparable to the baseline.
On the generative models (GPT-4.1, Claude, etc.), our approach typically consumes more tokens than the two baselines per successful run.
%
%Typically, the number of input tokens is two to three times of that of baseline, and the output tokens cost much more.
%
%The reason of token cost is straightforward.
%
This is a direct consequence of the design.
Our  method essentially substitutes the computational cost of the symbolic methods used in the baselines with the token cost of generating a detailed reasoning trace from the LLM.
On simpler problems, where the baselines are sufficient to solve, this can result in a higher token cost for our method.
%However, this verbosity is exactly the mechanism that enables our targeted feedback and leads to the superior effectiveness shown in RQ1.
%
However, the feedback mechanism of our approach is exactly what enables \toolname{} to solve more complex problems where the baseline approaches fall short, as shown in RQ1.
Note that we set the same resource limit for all methods in the evaluation.
The fact that \toolname{} can solve more difficult benchmarks within the same resource budget demonstrates that its precise feedback mechanism enables the LLM to discover non-trivial solutions that are inaccessible to the baselines.
%
%This indicates that by the precise feedback mechanism of our approach, the LLM has the ability to find non-trivial solutions in more difficult problems. 
%
Furthermore, despite the higher token count, the monetary cost remains acceptable.
For instance, a successful solve with GPT-4.1 costs approximately \$0.038 on average.

Regarding time costs, \clausetoinv{} is generally the most efficient.
This is because generating atomic clauses is a much simpler task than synthesizing complete invariants, resulting in lower time cost when solving easy problems.
However, it fails to solve more complex problems within the same 600 seconds limit compared to our approach.
For our approach and \baselinename{}, the time costs are generally comparable, with no approach having a consistent advantage across all models.
The runtime of our approach is dominated by LLM inference time, while \baselinename{}'s runtime is a mix of LLM inference and the computational cost of BMC.
It is interesting to note that when using the reasoning model GPT-o4-mini, \toolname{} is the most efficient in term of token cost among the three methods, and the time cost is much lower than \baselinename{}.
For the reasoning model, it is much more costly for the model to generate one response.
%
%For the reasoning model, the output token is much more than the input token. This is because the internal model reasoning process also counts tokens.
%
\baselinename{}'s reliance on multiple iterations leads to an average time of 227.9s.
\clausetoinv{} also requires multiple iterations to get a complete loop invariant, leading to a higher output token cost.
In contrast, \toolname{}'s ability to guide the model to a solution in fewer iterations (as noted in RQ1) results in a 2.5x time reduction than \baselinename{} and lower token usage, demonstrating the efficiency for models that excel at reasoning but may be slower to respond.

\subsubsection{RQ3: Impact of Feedback}
\label{sec:RQ3}

\begin{table}[t]
\centering
\caption{Classification of solved problems on the main benchmark suite. Feedback-Enhanced is a subset of Directly Solvable where feedback further improved consistency across runs.}
\label{tab:feedback_problem}
\begin{tabular}{l|ccc}
\toprule
\textbf{Model} & \textbf{Directly Solvable} & \textbf{Feedback-Enhanced} & \textbf{Feedback-Exclusive} \\
\midrule
GPT-4.1           & 305 & 112 & 140 \\
GPT-4.1-mini      & 258 & 103 & 181 \\
GPT-4o-mini       & 242 & 195 & 199 \\
GPT-o4-mini       & 324 & 316 & 111 \\
Claude-3.7-Sonnet & 322 & 199 & 121 \\
\bottomrule
\end{tabular}
\end{table}

\begin{table}[t]
\centering
\caption{Analysis of feedback contribution to successful runs on the main benchmark suite. A ``Direct Success'' occurs on the first attempt, while a ``Feedback-Driven Success'' requires one or more feedback iterations.}
\label{tab:feedback_run}
\begin{tabular}{l|ccc}
\toprule
\textbf{Model} & \textbf{Total Success} & \textbf{Direct Success} & \textbf{Feedback-Driven Success} \\
\midrule
GPT-4.1 & 2141 & 1259 (58.8\%) & 882 (41.2\%) \\
GPT-4.1-mini & 2095 & 1028 (49.1\%) & 1067 (50.9\%) \\
GPT-4o-mini & 1884 & 722 (38.3\%) & 1162 (61.7\%) \\
GPT-o4-mini & 2052 & 700 (34.1\%) & 1352 (65.9\%) \\
Claude-3.7-Sonnet & 2067 & 1089 (52.7\%) & 978 (47.3\%) \\
\bottomrule
\end{tabular}
\end{table}

\begin{table}[t]
\centering
\caption{The refinement success rate results on the main benchmark suite. The refinement success rate means the percentage of feedback iterations that resulted in an invariant with higher Jaccard similarity to the correct answer.}
\label{tab:refinement_success}
\begin{tabular}{l|c}
\toprule
\textbf{Model}  & \textbf{Refinement Success Rate} \\
\midrule
GPT-4.1  & 70.4\% \\
GPT-4.1-mini  & 71.4\% \\
GPT-4o-mini  & 62.9\% \\
GPT-o4-mini  & 79.6\% \\
Claude-3.7-Sonnet  & 80.2\% \\\midrule
\textbf{Average} & \textbf{72.9\%}  \\
\bottomrule
\end{tabular}
\end{table}

For some problems, the LLM can produce correct invariants directly from the initial prompt. Therefore, to isolate the specific contribution of our feedback mechanism, we analyze how many problems are actually solved by our feedback mechanism.
Table~\ref{tab:feedback_problem} partitions the solved problems into those solved on the initial attempt versus those that require feedback on the main benchmark suite.
Specifically, we classify the problems into the following three categories:

\begin{enumerate}
    \item \textit{Directly Solvable}: Problems where the LLM produces a correct loop invariant with the initial prompt in at least one run.
    \item \textit{Feedback-Exclusive}: Problems where the LLM fails to produce a correct loop invariant with the initial prompt in all five runs, but are successfully solved via our feedback mechanism. Feedback is the only path to a solution for these problems.
    \item \textit{Feedback-Enhanced}: A subset of the ``Directly Solvable'' category, where the LLM succeeds with the initial prompt in some runs but not all, and our feedback mechanism successfully repairs the initial failures in more runs. These problems can be solved occasionally without feedback but the success consistency can be greatly boosted using feedback.
\end{enumerate}

The data in Table~\ref{tab:feedback_problem} demonstrates that our feedback mechanism significantly improves model performance.
For GPT-4o-mini, 199 problems are solved exclusively by the feedback mechanism, reaching an 82.2\% increase over the model's direct solving capability.
For more powerful models like Claude-3.7-Sonnet, the feedback mechanism still yields a 37.6\% improvement.
Besides, our feedback mechanism narrows  the gap between less capable models and the state-of-the-art ones.
For instance, while the total number of problems solved by GPT-4.1-mini is only 6 fewer than GPT-4.1 using our approach (439 vs. 445), the number of problems solved \textit{directly} by GPT-4.1-mini is 47 fewer than GPT-4.1. In summary, our approach makes strong models stronger to solve more complex problems, and brings the performance of weaker models much closer to the state-of-the-art ones.

Beyond expanding capability, the feedback mechanism significantly improves stability. Since LLM generation is stochastic, a model may not succeed in all five runs directly for a given problem. The ``Feedback-Enhanced'' column in Table~\ref{tab:feedback_problem} lists the number of problems that can be solved directly, but where the success rate is improved by our feedback mechanism.
The results indicate that a large proportion of ``Directly Solvable'' problems benefit from refinement. For example, on GPT-o4-mini, 316 of the 324 direct-solve problems are improved by feedback in at least one run.
Table~\ref{tab:feedback_run} further breaks down all successful runs into direct successes and feedback-driven successes.
For GPT-o4-mini, 65.9\% of the total successful runs are achieved via feedback.
These results confirm that our approach stabilizes LLM performance, ensuring reliability on problems the model can theoretically solve on its own.

Finally, we inspect the quality of the refinement steps to determine how often the feedback leads to a better invariant.
We measure the refinement success rate in Table~\ref{tab:refinement_success}, defined as the percentage of feedback iterations where the refined invariant is closer to the correct answer than the previous proposal.
Specifically, we treat a loop invariant as the conjunction of a set of atomic clauses and calculate the Jaccard similarity ($J(A, B)=|A\cap B|/|A\cup B|$ for two sets $A$ and $B$) between the proposed invariant and the correct answer.
Clause matching is performed syntactically: two clauses are considered identical if they are identical after normalization (e.g., whitespace and operator canonicalization).
%For the correct answer, we use the final invariant produced by our tool in the currect run if the run eventually succeeds; otherwise, we use the ground truth.
The higher the Jaccard similarity is, the closer the proposed loop invariant is to the correct answer.
%A loop invariant is a conjunction of a set of atomic clauses. If the Jarccard similarity (defined as $|A\cap B|/|A\cup B|$ for two sets $A$ and $B$) between the proposed loop invariant and the correct answer is higher, we say the proposed loop invariant is better.

As shown in the table,  our feedback is highly constructive. The average success rate reaches 72.9\%. For GPT-4.1, 70.4\% of the refinement iterations result in an invariant closer to the correct answer. This rate is even higher for Claude-3.7-Sonnet and o4-mini. This high success rate confirms that the formal verification feedback provides constructive signals for the LLM toward the correct solution efficiently.

\subsubsection{RQ4: Combining with Symbolic Methods}

The motivation of our approach is to guide the LLM to generate a complete correct answer in the ``guess-and-check'' paradigm using constructive feedback.
%
%However, in the loop invariant synthesis problem, the LLM often generates individual correct loop invariants in multiple responses separately, which suggests the potential to combine multiple answers to get a new correct answer.
However, in the loop invariant synthesis problem, LLMs often produce partially correct conjuncts across several attempts.
\baselinename{}'s BMC-based combination strategy is designed to exploit this behavior.
%In the baseline method \baselinename{}, it applies bounded model checking (BMC) to filter candidate invariants after each iteration with the LLM and form a new candidate invariant.
%
Since this technique is orthogonal to our feedback loop, it is natural to combine them.
Specifically, after each feedback round, we use BMC to collect all valid conjuncts from the LLM's new response and add them to a growing set of answer set, which can serve as a new candidate invariant.
%Specifically, after each feedback round, we get a set of new candidates.
%
%Then, we use BMC to check the correctness of each candidate.
%
%For a conjunction, we split it into separate candidates.
%
%Thus, we get a set of candidates that not fail in the BMC.
%
%As the iteration grows, this set of candidates also grows, which approximates to the complete set.
%

\begin{table}[t]
\centering
\caption{Effect of combining \toolname{} with Bounded Model Checking (BMC). The solved runs by BMC are included in those by feedback.}
\label{tab:bmc}
\resizebox{\textwidth}{!}{%
\begin{tabular}{l|ccccc}
\toprule
\textbf{Method} & \textbf{Success Rate (\%)} & \textbf{\# Solved (Feedback)} & \textbf{\# Solved (BMC)} & \textbf{Tokens (I/O)} & \textbf{Time (s)} \\
\midrule
GPT-4.1 & 93.1 & 882 & 0 & 8296 / 2738 & 55.9 \\
GPT-4.1 + BMC & \textbf{93.6} & 921 & 162 & \textbf{7929 / 2643} & 50.7 \\
\midrule
GPT-4.1-mini & 91.1 & 1067 & 0 & 10203 / 3923 & 52.8 \\
GPT-4.1-mini + BMC & \textbf{91.6} & 1059 & 283 & \textbf{7525 / 2930} & 55.3 \\
\bottomrule
\end{tabular}
}
\end{table}

%\begin{table}[t]
 %   \centering
 %   \caption{Experiment Results that combine BMC.}
 %   \begin{tabular}{c|c|c|c|c|c|c}
 %   \hline
   %      Method & Success Rate (\%) & Total Directly & Total Feedback & BMC & Token (I/O) & Time (s) \\\hline\hline
  %       GPT-4.1 & 93.1 & 1259 & 882 & 0 & 8296/2738 & 55.9 \\
  %       GPT-4.1+BMC & 93.6 & 1233 & 921 & 162 & 7929/2643 & 50.7 \\\hline
  %       GPT-4.1-mini & 91.1 & 1028 & 1067 & 0 & 10203/3923 & 52.8 \\
   %      GPT-4.1-mini+BMC & 91.6 & 1047 & 1059 & 283 & 7525/2930 & 55.3 \\\hline
   % \end{tabular}
    
   % \label{tab:bmc}
%\end{table}

Table~\ref{tab:bmc} shows the results of this experiment on the main benchmark suite. The combination yields a marginal improvement in the overall success rate (e.g., +0.5\% for GPT-4.1). The more significant benefit is in efficiency. For GPT-4.1-mini, the hybrid approach reduces the input and output tokens by approximately 26\% and 25\%, respectively.

This outcome indicates that while our targeted feedback is powerful enough to guide the LLM to the full solution eventually, BMC can accelerate convergence by assembling the final invariant from partial pieces generated in earlier, less-successful iterations. This reduces the number of feedback loops required for some problems, thereby saving time and tokens. However, the small increase in total solved benchmarks suggests that for the most difficult problems, simply combining conjuncts is insufficient. The corrective guidance of our feedback mechanism remains essential for finding the missing, non-trivial parts of the invariant.

%We have studied how our approach combines with BMC for the model GPT-4.1 and GPT-4.1-mini.
%
%Table~\ref{tab:bmc} represents the experiment results.
%
%The study shows that by combining BMC, the overall success rate can only increment with a little, say, 0.5\%.
%
%The main improvement is on the token cost.
%
%For GPT-4.1-mini, combining with BMC leads to about 25\% save on the input and output tokens.
%
%This result shows that BMC can get a correct answer by combining multiple responses of the LLM within fewer feedback iterations.
%
%But only considering BMC cannot lead to more successful tries, the LLM will eventually get the correct answer through our detailed feedback.

\section{Discussion}
\label{sec:discuss}
In this section, we analyze the limitations and reliability of our proposed framework, and how our approach can generalize to more complex scenarios. Following the experimental results in Section~\ref{sec:evaluation}, we first conduct a detailed failure analysis to identify the root causes preventing the successful verification of the remaining benchmarks (Section~\ref{sec:failure_analysis}). Next, we discuss the correctness of the auto-formalization component through a manual validation (Section~\ref{sec:autoformalization_correctness}). Then, we provide guidelines for generalizing our approach to more complex programs with nested or multiple loops and memory manipulation (Section~\ref{sec:generalizing}). Finally, we discuss the broader threats to validity regarding our experimental design and benchmark selection (Section~\ref{sec:threats}).

\subsection{Failure Analysis}
\label{sec:failure_analysis}

For the  benchmarks that our method failed to solve in the evaluation of Section~\ref{sec:evaluation}, we conducted a manual inspection to better understand the limitations of our approach and get clues for the future work.
We found that a small subset of failures resulted from tool-specific limitations, such as Frama-C's handling of specific C features (e.g., \texttt{unsigned int} casting), or the benchmark code itself cannot be verified.
For the remaining failures, the root causes fall into two primary categories: (1) the inability of the feedback to guide the LLM's global reasoning, and (2) ineffective refinement strategies where the LLM weakens invariants rather than strengthening them to satisfy verification conditions.
We illustrate these two failure modes via the following case studies.

\subsubsection{Failure on Global Reasoning}

The first failure category occurs when the problem requires understanding the ``big picture'', for example, identifying dead code or global constraints that local reasoning steps miss.
Our feedback mechanism focuses on local logical errors of the LLM's thinking process, which helps reduce hallucination, but it is less effective when the LLM's fundamental understanding of the control flow is incorrect.

\begin{figure}[t]
    \centering

    \begin{minipage}{0.44\textwidth}

    \begin{lstlisting}[language=C,basicstyle=\ttfamily\small]
//Pre-condition: the values of variable x, y, w, z are all 0.
while(unknown()) {
    if(unknown()) {
        x++; y += 100;
    } else if(unknown()) {
        if(x >= 4) {
            x++; y++;
        }
    } else if (y > 10 * w) {
        if(z >= 100 * x) 
            y = -y;
    }
    w++;
    z += 10;
}

assert((x < 4) || (y > 2));
    \end{lstlisting}
    \caption{Example program where our approach failed on global reasoning.}
    \label{fig:failure-case1}

    \end{minipage}
\hfill
    \begin{minipage}{0.52\textwidth}
\vspace{-0.2in}
    \begin{minipage}{\linewidth}
    \begin{lstlisting}[language=C,basicstyle=\ttfamily\small]
/*@
loop invariant i1: z == 10 * w;
loop invariant i2: y <= 100 * x;
loop invariant i3: (x < 4) ==> (y == 100 * x);
loop invariant i4: (x >= 4) ==> (y >= 400);
*/
    \end{lstlisting}
    \vspace{-0.2in}
    \caption{Correct loop invariants for the program.}
    \vspace{0.2in}
    \label{fig:correct-case1}
    \end{minipage}
%\vspace{2in}
    \begin{minipage}{\linewidth}
\begin{lstlisting}[language=C,basicstyle=\ttfamily\small]
/*@
loop invariant i1: x >= 0;
loop invariant i2: y >= 0 || ( y < 0 && z >= 100 * x);
loop invariant i3: z == 10 * w;
loop invariant i4: w >= 0;
loop invariant i5: (x < 4) || (y > -100 * x);
*/
    \end{lstlisting}
    \vspace{-0.2in}
    \caption{Loop invariants proposed by the LLM.}
    \label{fig:proposed-case1}
    \end{minipage}
    \end{minipage}

\end{figure}

Figure~\ref{fig:failure-case1} presents a program that our method failed to solve using GPT-4.1, and Figure~\ref{fig:correct-case1} shows the correct loop invariants needed to verify the assertion.
The loop contains three branches, and proving the assertion requires realizing the third branch (containing $y=-y$) is unreachable.
This is deduced from the global upper bound $y\leq 100*x$ and $z=10*w$. The branch condition requires $y>10 * w$, which implies $y>z$. However, since $y\leq 100*x$ and the branch also requires $z\geq 100*x$, this leads to a contradiction.
Because this branch is unreachable, the variable \textit{y} is never negated and remains non-negative, making the assertion provable.
However, the LLM failed to identify this unreachability. As shown in Figure~\ref{fig:proposed-case1}, it proposed invariants \textit{i2} and \textit{i5} based on the assumption that \textit{y} could be negative.
Once this incorrect assumption is made, the preservation must be verified across all three branches, which fails in this case.
After feedback, the LLM attempted to fix the logic for negative \textit{y} rather than realizing \textit{y} cannot be negative.
This highlights a limitation of our method: while our method effectively repairs local reasoning errors, it cannot currently prompt the LLM to re-think the global reasoning of the problem.
However, as LLMs become more and more powerful, our method will be more helpful to leverage their abilities, as shown in our evaluation.

\subsubsection{Ineffective Refinement}

The second failure category involves the LLM's inability to repair invariants when preservation fails. Instead of strengthening the invariant to complete the inductive proof, the LLM sometimes attempts to weaken the invariant or delete the invariant entirely.

\begin{figure}[t]
    \centering

    \begin{minipage}{0.44\textwidth}

    \begin{lstlisting}[language=C,basicstyle=\ttfamily\small]
int main() {
    int j, k, flag;
    j = 2;
    k = 0;

    while(unknown()) {
        if(flag) 
            j = j + 4;
        else {
            j = j + 2;
            k = k + 1;
        }
    }

    assert((k == 0) || j == 2 * k + 2);
}
    \end{lstlisting}
    \caption{Example program where our approach failed by ineffective refinement.}
    \label{fig:failure-case2}

    \end{minipage}
\hfill
    \begin{minipage}{0.52\textwidth}
    \vspace{-0.17in}
    \begin{minipage}{\linewidth}
    \begin{lstlisting}[language=C,basicstyle=\ttfamily\small]
/*@
loop invariant i1: k >= 0;
loop invariant i2: j >= 2;
loop invariant i3: (flag == 0) ==> (j == 2 * k + 2);
loop invariant i4: (flag != 0) ==> (k == 0);
*/
    \end{lstlisting}
    \vspace{-0.2in}
    \caption{Correct loop invariants for the program.}
    \vspace{0.2in}
    \label{fig:correct-case2}
    \end{minipage}
%\vspace{2in}
    \begin{minipage}{\linewidth}
\begin{lstlisting}[language=C,basicstyle=\ttfamily\small]
/*@
loop invariant i1: k >= 0;
loop invariant i2: j >= 2;
loop invariant i3: j == 2 * k + 2 || k == 0;
*/
    \end{lstlisting}
    \caption{Loop invariants proposed by the LLM.}
    \label{fig:proposed-case2}
    \end{minipage}
    \end{minipage}

\end{figure}

Figure~\ref{fig:failure-case2} illustrates this scenario. The program execution follows two distinct traces determined by \textit{flag}: if \textit{flag} is set, $k$ remains 0; otherwise, $j$ and $k$ maintain a linear relationship.
The LLM correctly identified the two states but combined them into a single disjunctive invariant: $(j=2*k+2)\lor(k=0)$ (Figure~\ref{fig:proposed-case2}).
While this property holds, it is not inductive on its own.
To prove preservation, the verifier must know \textit{which} branch of the disjunction holds to determine the valid transition. 
For example, if $j=2*k+2$, executing the \textit{if(flag)} branch breaks the relationship unless we know \textit{flag} prevents the linear relation trace.
The correct solution requires conditioning the invariant on \textit{flag} (see Figure~\ref{fig:correct-case2}).

When our system provided feedback that preservation failed, the LLM did not introduce \textit{flag} to split the cases. Instead, it attempted to weaken the invariant to a congruence modulo 4 (e.g., $(j - 2 * k - 2) \% 4 = 0$), or simply deleted the invariant.
While the weakened invariant is preserved, it is too weak to prove the final assertion.
This suggests that future work should explore guiding the LLM to refine the loop invariants for inductive verification.

\subsection{Correctness of Auto-Formalization}
\label{sec:autoformalization_correctness}

A core component of our framework is the auto-formalization of the Synthesizer LLM's natural language proof into first-order logic implications. A significant concern is the potential for the Formalizer LLM to hallucinate or misinterpret the natural language reasoning. If the translation is unfaithful, the resulting feedback could pinpoint non-existent errors or miss actual flaws.
To mitigate this risk, we have implemented several design strategies.
First, instead of requesting a complete proof script (e.g. in Coq or Isabelle), which necessitates strict adherence to complex syntax and global context, we instruct the LLM to translate individual reasoning steps into first-order logic implications.
This task is much simpler than generating a full proof script and is well-suited for our purpose of detecting local reasoning errors in the LLM's thinking process.
Second, we prompt the Synthesizer LLM to generate the natural language proof step-by-step, allowing the Formalizer LLM to focus on one specific step at a time.
This decomposition reduces the problem size, thereby enhancing the precision of the formalization procedure.
We also carefully design the prompts to describe the task accurately.

However, this procedure cannot always be perfect, and the LLM can still make mistakes or hallucinate during formalization.
To empirically validate the correctness of this procedure, we conduct a manual evaluation on a random sample of ten benchmark runs from our evaluation suite using GPT-4.1.
For each run, we manually inspect every generated formalization step to verify whether the logical implications generated by the LLM accurately reflect the meaning of the corresponding natural language proof of the step.
The results of this inspection are summarized in Table~\ref{tab:formalization_accuracy}. Across the sampled runs, the framework generates a total of 108 formalization steps. Our manual review finds that 90 of these steps (83.3\%) are correctly formalized.

\begin{table}[t]
\centering
\caption{Manual evaluation of auto-formalization correctness on 108 sampled proof steps.}
\label{tab:formalization_accuracy}
\small
\begin{tabular}{lrr}
\toprule
\textbf{Benchmark} & \textbf{Total Steps} & \textbf{Correct Steps} \\
\midrule
cav/gulv\_simp & 7 & 5 \\
pie/hola/10 & 15 & 12 \\
loop-lit/cggmp2005 & 10 & 9 \\
loop-zilu/benchmark15\_conjunctive & 11 & 11 \\
dagger/cars & 11 & 9 \\
code2inv/29 & 14 & 9 \\
llreve/barthe\_merged\_safe & 7 & 7 \\
loop-acceleration/diamond\_2\_2 & 7 & 6 \\
SyGuS/153 & 17 & 15 \\
loop-invariants/eq1-2 & 9 & 7 \\
\midrule
\textbf{Total} & \textbf{108} & \textbf{90 (83.3\%)} \\
\bottomrule
\end{tabular}
\end{table}

We also analyze the incorrect formalizations to understand the causes of failure. 
We observe that the LLM performs well with pure mathematical reasoning but struggles with specific structural or semantic translations.
The primary sources of error were:
\begin{enumerate}
    \item \textit{Case Analysis:} Challenges arise during case analysis. In some instances, a natural language step merely states a case division (e.g., claiming the domain can be divided into $x>4$ and $x \le 4$). However, the LLM occasionally repeats the overall proof goal within these case definitions, rendering the implication invalid.
    \item \textit{Program Semantics:} The LLM sometimes fails to distinguish between program state updates and logical equality. For example, it may formalize a variable update as $x=x+1$. While valid in C syntax, this is a contradiction in standard mathematical logic. Although we explicitly address this in the prompt, the model occasionally remains confused.
    \item \textit{Context Omission:} The Formalizer LLM occasionally fails to include implicit pre-conditions mentioned in the natural language (e.g., omitting a variable bound previously established). This leads to implications that are logically invalid despite being semantically aligned with the text.
\end{enumerate}

Errors by the Formalization LLM can introduce ``noise'' into the feedback mechanism.
If the Formalizer LLM produces an invalid implication from a correct natural language step (a false positive error detection), the resulting feedback may pinpoint an error where the Synthesizer LLM is actually correct.
On the contrast, if the Formalizer LLM incorrectly validates a flawed step (a false negative), the error in reasoning is missed.
Additionally, our formalization does not verify structural errors in the proof.
We employ specific strategies to mitigate the impact of this noise. In the case of false positives, we choose to report all errors found in the formalized proof. While this may include spurious feedback, it ensures that genuine reasoning errors are not suppressed. In the case of false negatives, even if a specific reasoning step is missed, the feedback mechanism can at least identify the failure at the level of the specific verification condition (e.g., preservation failure).
 While mistakes in auto-formalization cannot be entirely eliminated, their impact on the overall framework is limited to efficiency rather than soundness. 
Since the final acceptance of loop invariants is strictly guarded by the formal verification tool (Frama-C), LORIS will never accept an incorrect solution based on faulty auto-formalization. The worst-case scenario is a delay in convergence, not a verification failure. The high accuracy rate of the sampled auto-formalization (83.3\%) and the strong overall results presented in Section~\ref{sec:evaluation} suggest that for the majority of iterations, the feedback provided is accurate and constructive.

\subsection{Generalizing to Complex Scenarios}
\label{sec:generalizing}

While the current implementation of \toolname{} focuses on numerical problems with single loops to establish a fair comparison with state-of-the-art baselines, the underlying methodology can generalize to more complex programs. 
The core principle - guiding an LLM by formally verifying its local reasoning - is generic and applicable to more sophisticated verification scenarios.
In this section, we discuss how our approach can generalize to real-world C programs. The generalization includes two main directions: nested or multiple loops and complex memory logic.

\subsubsection{Nested or Multiple Loops}

As standard verification tools like Frama-C have the ability to verify programs with nested or multiple loops through their loop invariants, our approach can scale to those programs seamlessly. Specifically, for a program with nested or multiple loops, we prompt the LLM to propose loop invariants for all loops simultaneously. Frama-C will generate the establishment and preservation verification conditions (VCs) for each loop invariant. Similarly to  the approach in Section~\ref{sec:approach-feedback}, we select a single failed VC and let the LLM express the idea to prove it in natural language. The order for selecting the specific failed VC still follows the natural deductive order: from prior loops to posterior loops, and from inner loops to outer loops. Then the natural language proof is formalized and checked, and the feedback is constructed. 
The refinement will require propagating the error to other loops, since the proofs for the invariants of multiple loops are dependent with each other.
For example, proving the establishment of an inner loop invariant often requires strengthening the invariant of the enclosing outer loop.

To illustrate this, we provide a case study. Figure~\ref{fig:program-nestedloop} presents a program with a nested loop. This code implements the Extended Euclidean Algorithm, which finds the greatest common divisor (GCD) of two numbers $x$ and $y$.  The assertion to prove is the linear combination $p*x+r*y=a$. This problem represents a challenge of nested loops because the correctness of the inner loop depends on the bounds maintained by the outer loop.

We chat with Google's Gemini-3-Flash, and the LLM produces the loop invariants for the outer loop and the inner loop in Figure~\ref{fig:outer-loop} and Figure~\ref{fig:inner-loop}. The LLM correctly identifies the linear relations for both the outer loop and the inner loop. However, the verification fails on the establishment of the inner loop invariant $c\ge 0$. 
While semantically correct, Frama-C cannot verify this property because the necessary context that $a$ (assigned to $c$ at line 9) is non-negative is missing from the outer loop's invariant.

We then prompt the LLM to prove $c\ge 0$ holds at the start of the inner loop. In the natural language proof, the LLM erroneously relies on the initial program state ($c=0$ at line 6) rather than the state at the start of the current outer loop iteration. This condition ($c=0$) is then listed as the initial condition in the formalized proof, and can be verified invalid by Frama-C following the approach in Section~\ref{sec:approach-feedback}. After the feedback, the LLM understands that it misuses the precondition, and strengthens the outer loop invariants to include $a\ge 0$ and $b\geq 0$, enabling the successful verification of the entire program. This case study demonstrates that our local reasoning feedback can scale to programs with nested loops.

\subsubsection{Heap Manipulating}

The other direction of generalization is scaling beyond numerical domains to programs involving complex data structures and memory manipulation. This requires extending the underlying logical framework to Separation Logic.

To adapt our framework for broader logical theories, the pipeline requires two modular updates. First, the Formalizer LLM must be prompted to support theories specific to memory reasoning, such as spatial conjunctions for heap representations. The recent work~\cite{LiuWFCY24} has demonstrated that LLMs possess a foundational understanding of Separation Logic and can generate valid Separation Logic assertions. Second, the verification backend must be substituted with a solver capable of handling these theories, such as CVC5~\cite{cvc5} or SongBird~\cite{TaLKC19}. With these modifications, the core mechanism of \toolname{} remains unchanged: the Synthesizer LLM proposes memory invariants, and the feedback loop corrects local reasoning errors regarding heap validity. As LLMs perform poorly on generating memory loop invariants directly~\cite{LiuWFCY24}, our approach has the potential to improve it.

\begin{figure}[t]
    \centering

    \begin{minipage}[t]{0.44\textwidth}

    \begin{lstlisting}[language = C ,  basicstyle=\ttfamily\small, numbers=left,firstnumber = 1 , escapeinside={(*@}{@*)}]
int main() {
    int x, y;
    int a, b, p, q, r, s, c, k;
    a = x; b = y;
    p = s = 1;
    q = r = c = k = 0;
    
    while (b != 0) {
        c = a;
        k = 0;
        while (c >= b) {
            c = c - b;
            k = k + 1;
        }
        \end{lstlisting}
        \end{minipage}
        \quad\quad
        \begin{minipage}[t]{0.44\textwidth}
\begin{lstlisting}[language = C ,  basicstyle=\small\ttfamily, numbers=left,firstnumber = last, escapeinside={(*@}{@*)}]
        a = b;
        b = c;

        long long temp;
        temp = p;
        p = q;
        q = temp - q * k;
        temp = r;
        r = s;
        s = temp - s * k;
    }
    assert (p * x + r * y == a); 
    return a;
}
    \end{lstlisting}
    \end{minipage}
    \caption{Example program with nested loops.}
    \label{fig:program-nestedloop}
    \end{figure}
\begin{figure}[t]
    \begin{minipage}[t]{0.48\textwidth}
    \begin{lstlisting}[language=C,basicstyle=\ttfamily\small]
/*@ 
loop invariant a == p * x + r * y;
loop invariant b == q * x + s * y; 
*/
    \end{lstlisting}
    \caption{Loop invariants for the outer loop proposed by the LLM.}
    \label{fig:outer-loop}
    \end{minipage}
%\vspace{2in}
\quad
    \begin{minipage}[t]{0.48\textwidth}
\begin{lstlisting}[language=C,basicstyle=\ttfamily\small]
/*@
loop invariant c >= 0;
loop invariant a == k * b + c;
*/
    \end{lstlisting}
    \caption{Loop invariants for the inner loop proposed by the LLM.}
    \label{fig:inner-loop}
    \end{minipage}

\end{figure}

\subsection{Threats to Validity}
\label{sec:threats}

We identify several threats to the validity of our study and the limitations of our current approach.

\paragraph{External Validity.} 
The primary threat to external validity concerns the generalizability of our benchmarks. 
Our evaluation datasets consists of a main benchmark suite of 460 C programs and a non-linear benchmark  suite of 50 C programs. While this suite covers a diverse range of linear and non-linear numerical properties, the programs are limited to single-loop functions involving numerical arithmetic. 
The benchmark does not contain programs involving complex data structures such as arrays, pointers, or recursive data types due to the limited capabilities of SMT solver.
%We explicitly excluded programs requiring floating-point arithmetic  and those involving complex data structures such as arrays, pointers, or recursive data types. 
Consequently, while LORIS demonstrates high effectiveness on numerical invariants, its performance on real-world software systems involving heap manipulation or complex memory models remains unverified. 
However, we posit that the proposed feedback mechanism—identifying and correcting local reasoning errors via auto-formalization—is generally applicable to these more complex scenarios.  The feedback mechanism is agnostic to the specific program domain. While the current pipeline is constrained by the solver, the methodology of guiding LLMs via formal verification feedback is expected to generalize to broader classes of programs. We discuss how to generate our approach to more complex scenarios like nested or multiple loops and more complex logic in Section~\ref{sec:generalizing}.

\paragraph{Construct Validity.} 
A core component of our framework is the \textit{auto-formalization} step, where an LLM translates natural language reasoning into first-order logic implications. 
A threat to construct validity is the fidelity of this translation. If the Formalizer LLM hallucinates or misinterprets the natural language step, the resulting feedback to the Synthesizer LLM may be misleading. 
Although we mitigate this by instructing the model to perform a lightweight, literal translation rather than generating complex proof scripts, the translation process is not formally verified. 
However, it is important to note that this threat affects only the \textit{efficiency} of the synthesis loop, not the \textit{soundness} of the final result. 
Because the final invariants are always verified by Frama-C, incorrect feedback might delay convergence or lead to a timeout, but it will not result in the acceptance of an incorrect invariant.
Besides, we have validated that the accuracy of the auto-formalization step through a manual inspection in Section~\ref{sec:autoformalization_correctness}.

\paragraph{Internal Validity.} 
To mitigate the stochastic nature of LLMs, we performed five independent runs for each benchmark and reported the aggregated results. 
In terms of baselines, we compared LORIS with \textsc{Lam4Inv}  and \textsc{Clause2Inv}. 
While  these works represent the state-of-the-art in LLM-assisted loop invariant synthesis, the rapid evolution of this field means newer techniques may exist. 
Additionally, our reliance on Frama-C  and Z3  implies that our tool's performance is upper-bounded by the capabilities of these underlying solvers. 
As noted in Section~\ref{sec:failure_analysis}, some failures were attributed to Frama-C's handling of specific C features rather than a failure of our logic.

\section{Related Work}
\label{sec:related}

Our work integrates Large Language Models with formal verification feedback on the model's thinking process to synthesize loop invariants.
It is related to several research areas, including loop invariant synthesis, the application of LLMs to synthesize loop invariants, and auto-formalization.

\paragraph{Traditional Loop Invariant Synthesis}

The automatic synthesis of loop invariants is a long-standing challenge in program analysis and verification.
Traditional approaches largely follow the ``guess-and-check'' paradigm, in which a loop invariant is first proposed and then checked for validity.
Symbolic methods, such as those based on abstract interpretation~\cite{BlanchetCCFMMMR03,CousotC77} and predicate abstraction~\cite{JhalaM06,Mcmillan10}, systematically explore a state space defined by abstract domains or pre-defined predicates.
Template-based methods~\cite{SomenziB11,ErnstCGN00} presuppose a parametric form for invariants and use solvers to find suitable parameter values.
While effective, these methods are often constrained by the expressiveness of their pre-defined templates or abstract domains.
Data-driven approaches, on the other hand, infer invariants from program execution traces~\cite{ErnstCGK99,PadhiSM16}, or use machine learning models~\cite{Code2Inv18,CLI2INV19,YaoRWJG20,YuWW23} trained on large datasets of programs and their invariants.
These methods can discover complex invariants but typically require substantial and domain-specific training data.
In contrast, our approach leverages a pre-trained LLM, which does not require task-specific training and is not confined to rigid templates, allowing for the generation of more diverse and intuitive candidate invariants without heavy feature-engineering.

\paragraph{LLMs for Loop Invariant Synthesis}

With their remarkable success in code generation and understanding, LLMs have recently been applied to various programming tasks.
Several studies have explored using LLMs for loop invariant generation.
\citet{AdharshKSLDRR23} directly leverage LLMs to generate inductive loop invariants and provides the LLM with a pass/fail feedback.
\citet{WenCSXQHLCT24} propose an approach that synthesizes program specifications in addition to loop invariants for verification.
\citet{wu2024lemur} employ LLMs to generate program properties that help the verification task.
LaM4Inv~\cite{LaM4Inv} uses bounded model checking to filter correct invariants from multiple responses of the LLM, and provides counterexamples generated by an SMT solver when an invariant fails verification as feedback.
Clause2Inv~\cite{Clause2Inv} prompts the LLM to generate atomic clauses rather than complete loop invariants, and then logically combine them using a heuristic-based algorithm.
It also provides counterexamples as feedback to the LLM.
Compared to previous works, which give a simple pass/fail or counterexample feedback, our approach goes further by analyzing the LLM's own thinking process to explain why it is wrong.
By identifying the specific logical flaw in the LLM's thinking process, we provide a more fundamental and constructive form of feedback.

\paragraph{Auto-formalization by LLMs}

A key component of our framework is the formalization of the LLM's natural language proof.
This relates to the field of auto-formalization, which aims to translate informal, natural language text into formal, machine-readable logic.
Recent efforts have explored using LLMs to generate formal proof scripts for interactive theorem provers like Coq~\cite{Coq97}, Isabelle~\cite{Isabelle02}, or Lean~\cite{Lean15}.
\citet{jiang2023draft} develop a ``Draft, Sketch, and Prove'' framework, which maps informal proofs to formal proof sketches using LLMs, and then uses the sketches to guide an automated prover to search for proofs for easier sub-problems.
\citet{zhou2024dont} solve mathematical quantitative reasoning problems by transforming the problem statement and solution into Isabelle, and let an LLM give the formal proof for checking.
\citet{LuDZ24} first prompt an LLM to generate an initial Coq proof, and then leverage targeted symbolic methods to repair the low-level problems.
Compared to the auto-formalization works, the purpose of our approach is different.
Our work does not aim to generate a complete, verifiable proof script as the end product.
Instead, we guide the LLM towards the correct solution in the ``guess-and-check'' problem by pointing out the logical error in its thinking process.
This leads to the lightweight formalization which can be checked by an SMT solver.

\section{Conclusion and Future Work}
\label{sec:conclusion}

In this paper, we addressed a fundamental challenge in applying Large Language Models to ``guess-and-check'' problems: while LLMs excel at generating high-level, intuitive solutions, they often fail to ensure their correctness due to hallucinations and logical errors. We proposed a novel paradigm that guides an LLM to refine its own solution by providing formal verification feedback on the local and detailed errors in its thinking process.

We instantiated this paradigm in a framework called \toolname{} for synthesizing loop invariants in C programs. \toolname{} prompts the LLM not only for a candidate invariant but also for a step-by-step natural language proof of its correctness. This proof is then formalized into a series of lightweight, first-order logic implications, which are then checked by an SMT solver to pinpoint the exact logical flaws. These flaws are then presented to the LLM as precise, constructive feedback for refinement. Our evaluation on a main benchmark suite of 460 programs and an additional benchmark suite of 50 programs each of which involves non-linear properties demonstrates the effectiveness of this approach. Using GPT-4.1, \toolname{} solved 445 programs,  achieving a success rate of  93.1\% on the main benchmark suite. On the more challenging non-linear benchmark suite, our approach can solve 47 of the 50 programs.

Our work opens several promising avenues for future research. A natural next step is to extend \toolname{} to handle more complex verification challenges. This includes synthesizing invariants for programs with multiple or nested loops, which requires reasoning about the interplay between different invariants, as well as handling programs with complex data structures and heap manipulations, which would necessitate a more expressive logic for the invariants and proofs.

More broadly, the core feedback paradigm proposed in this paper is not limited to loop invariant synthesis. We believe it can be generalized to a wide range of ``guess-and-check'' problems where LLMs can provide an initial solution sketch. For instance, in automated theorem proving, our method could be used to ask an LLM to justify a difficult proof step, formalize its explanation, and identify the logical leap. 
%In program synthesis, when a generated program fails a test case, our method could prompt the LLM to trace its execution and then formally check this trace to find where the model's understanding of its own code is flawed. 
Exploring these and other domains would be a valuable direction to further validate and generalize the effectiveness of guiding LLMs through a formal critique of their own reasoning.

\section*{Data Availability Statement}
We release the replication package of \toolname{} on \url{https://github.com/ltcRandomwalk/LORIS}.
It includes all the source code, scripts, benchmarks and evaluation results.

\section*{Acknowledgments}
We would like to thank the anonymous reviewers for their constructive feedback. This work was supported in part by the National Natural Science Foundation of China under Grant  No. W2411051, and Ant Group Research Fund.

%% Acknowledgments
%\begin{acks}                            %% acks environment is optional
                                        %% contents suppressed with 'anonymous'
  %% Commands \grantsponsor{<sponsorID>}{<name>}{<url>} and
  %% \grantnum[<url>]{<sponsorID>}{<number>} should be used to
  %% acknowledge financial support and will be used by metadata
  %% extraction tools.
  %The work is supported by the \grantsponsor{GS100000001}{National Natural %Science Foundation of China}{http://dx.doi.org/10.13039/100000001}
%National Natural Science Foundation of China under Grant No. 62172017.
  %This material is based upon work supported by the
  %\grantsponsor{GS100000001}{National Science
   % Foundation}{http://dx.doi.org/10.13039/100000001} under Grant
  %No.~\grantnum{GS100000001}{nnnnnnn} and Grant
  %No.~\grantnum{GS100000001}{mmmmmmm}.  Any opinions, findings, and
 % conclusions or recommendations expressed in this material are those
  %of the author and do not necessarily reflect the views of the
 % National Science Foundation.
 
%Acknowledgments
%\end{acks}
%

%% Bibliography
\bibliographystyle{ACM-Reference-Format}
\bibliography{main,x1-LLM-loop}

@article{HuangYMZFZQW25,
author = {Huang, Lei and Yu, Weijiang and Ma, Weitao and Zhong, Weihong and Feng, Zhangyin and Wang, Haotian and Chen, Qianglong and Peng, Weihua and Feng, Xiaocheng and Qin, Bing and Liu, Ting},
title = {A Survey on Hallucination in Large Language Models: Principles, Taxonomy, Challenges, and Open Questions},
year = {2025},
issue_date = {March 2025},
publisher = {Association for Computing Machinery},
address = {New York, NY, USA},
volume = {43},
number = {2},
issn = {1046-8188},
url = {https://doi.org/10.1145/3703155},
doi = {10.1145/3703155},
journal = {ACM Trans. Inf. Syst.},
month = jan,
articleno = {42},
numpages = {55}
}

@inproceedings{SomenziB11,
author = {Somenzi, Fabio and Bradley, Aaron R.},
title = {IC3: where monolithic and incremental meet},
year = {2011},
isbn = {9780983567813},
publisher = {FMCAD Inc},
address = {Austin, Texas},
booktitle = {Proceedings of the International Conference on Formal Methods in Computer-Aided Design},
pages = {3–8},
numpages = {6},
location = {Austin, Texas},
series = {FMCAD '11}
}

@inproceedings{ErnstCGN00,
author = {Ernst, Michael D. and Czeisler, Adam and Griswold, William G. and Notkin, David},
title = {Quickly detecting relevant program invariants},
year = {2000},
isbn = {1581132069},
publisher = {Association for Computing Machinery},
address = {New York, NY, USA},
url = {https://doi.org/10.1145/337180.337240},
doi = {10.1145/337180.337240},
booktitle = {Proceedings of the 22nd International Conference on Software Engineering},
pages = {449–458},
numpages = {10},
location = {Limerick, Ireland},
series = {ICSE '00}
}

@InProceedings{SiNDNS20,
author="Si, Xujie
and Naik, Aaditya
and Dai, Hanjun
and Naik, Mayur
and Song, Le",
editor="Lahiri, Shuvendu K.
and Wang, Chao",
title="Code2Inv: A Deep Learning Framework for Program Verification",
booktitle="Computer Aided Verification",
year="2020",
publisher="Springer International Publishing",
address="Cham",
pages="151--164",
isbn="978-3-030-53291-8"
}

@misc{GabrielJJYJ19,
      title={CLN2INV: Learning Loop Invariants with Continuous Logic Networks}, 
      author={Gabriel Ryan and Justin Wong and Jianan Yao and Ronghui Gu and Suman Jana},
      year={2019},
      eprint={1909.11542},
      archivePrefix={arXiv},
      primaryClass={cs.LG},
      url={https://arxiv.org/abs/1909.11542}, 
}

@InProceedings{GargLMDN14,
author="Garg, Pranav
and L{\"o}ding, Christof
and Madhusudan, P.
and Neider, Daniel",
editor="Biere, Armin
and Bloem, Roderick",
title="ICE: A Robust Framework for Learning Invariants",
booktitle="Computer Aided Verification",
year="2014",
publisher="Springer International Publishing",
address="Cham",
pages="69--87",
isbn="978-3-319-08867-9"
}

@misc{AdharshKSLDRR23,
      title={Finding Inductive Loop Invariants using Large Language Models}, 
      author={Adharsh Kamath and Aditya Senthilnathan and Saikat Chakraborty and Pantazis Deligiannis and Shuvendu K. Lahiri and Akash Lal and Aseem Rastogi and Subhajit Roy and Rahul Sharma},
      year={2023},
      eprint={2311.07948},
      archivePrefix={arXiv},
      primaryClass={cs.PL},
      url={https://arxiv.org/abs/2311.07948}, 
}

@inproceedings{PeiBSY23,
author = {Pei, Kexin and Bieber, David and Shi, Kensen and Sutton, Charles and Yin, Pengcheng},
title = {Can large language models reason about program invariants?},
year = {2023},
publisher = {JMLR.org},
booktitle = {Proceedings of the 40th International Conference on Machine Learning},
articleno = {1144},
numpages = {25},
location = {Honolulu, Hawaii, USA},
series = {ICML'23}
}

@InProceedings{CuoqKPSSY12,
author="Cuoq, Pascal
and Kirchner, Florent
and Kosmatov, Nikolai
and Prevosto, Virgile
and Signoles, Julien
and Yakobowski, Boris",
editor="Eleftherakis, George
and Hinchey, Mike
and Holcombe, Mike",
title="Frama-C",
booktitle="Software Engineering and Formal Methods",
year="2012",
publisher="Springer Berlin Heidelberg",
address="Berlin, Heidelberg",
pages="233--247",
isbn="978-3-642-33826-7"
}

@inproceedings{LuDZ24,
author = {Lu, Minghai and Delaware, Benjamin and Zhang, Tianyi},
title = {Proof Automation with Large Language Models},
year = {2024},
isbn = {9798400712487},
publisher = {Association for Computing Machinery},
address = {New York, NY, USA},
url = {https://doi.org/10.1145/3691620.3695521},
doi = {10.1145/3691620.3695521},
booktitle = {Proceedings of the 39th IEEE/ACM International Conference on Automated Software Engineering},
pages = {1509–1520},
numpages = {12},
location = {Sacramento, CA, USA},
series = {ASE '24}
}

@article{Hoare1969,
author = {Hoare, C. A. R.},
title = {An axiomatic basis for computer programming},
year = {1969},
issue_date = {Oct. 1969},
publisher = {Association for Computing Machinery},
address = {New York, NY, USA},
volume = {12},
number = {10},
issn = {0001-0782},
url = {https://doi.org/10.1145/363235.363259},
doi = {10.1145/363235.363259},
journal = {Commun. ACM},
month = oct,
pages = {576–580},
numpages = {5}
}

@inproceedings{LaM4Inv,
author = {Wu, Guangyuan and Cao, Weining and Yao, Yuan and Wei, Hengfeng and Chen, Taolue and Ma, Xiaoxing},
title = {LLM Meets Bounded Model Checking: Neuro-symbolic Loop Invariant Inference},
year = {2024},
isbn = {9798400712487},
publisher = {Association for Computing Machinery},
address = {New York, NY, USA},
url = {https://doi.org/10.1145/3691620.3695014},
doi = {10.1145/3691620.3695014},
booktitle = {Proceedings of the 39th IEEE/ACM International Conference on Automated Software Engineering},
pages = {406–417},
numpages = {12},
location = {Sacramento, CA, USA},
series = {ASE '24}
}

@inproceedings{MadhukarWKLS15,
author = {Madhukar, Kumar and Wachter, Bj\"{o}rn and Kroening, Daniel and Lewis, Matt and Srivas, Mandayam},
title = {Accelerating invariant generation},
year = {2015},
isbn = {9780983567851},
publisher = {FMCAD Inc},
address = {Austin, Texas},
booktitle = {Proceedings of the 15th Conference on Formal Methods in Computer-Aided Design},
pages = {105–111},
numpages = {7},
location = {Austin, Texas},
series = {FMCAD '15}
}

@article{ZhuMJ18,
author = {Zhu, He and Magill, Stephen and Jagannathan, Suresh},
title = {A data-driven CHC solver},
year = {2018},
issue_date = {April 2018},
publisher = {Association for Computing Machinery},
address = {New York, NY, USA},
volume = {53},
number = {4},
issn = {0362-1340},
url = {https://doi.org/10.1145/3296979.3192416},
doi = {10.1145/3296979.3192416},
journal = {SIGPLAN Not.},
month = jun,
pages = {707–721},
numpages = {15}
}

@online{svcomp,
author={SVCOMP},
title={Competition on software verification},
year=2025,
url={https://sv-comp.sosy-lab.org}
}

@inproceedings{z3,
author = {De Moura, Leonardo and Bj\o{}rner, Nikolaj},
title = {Z3: an efficient SMT solver},
year = {2008},
isbn = {3540787992},
publisher = {Springer-Verlag},
address = {Berlin, Heidelberg},
booktitle = {Proceedings of the Theory and Practice of Software, 14th International Conference on Tools and Algorithms for the Construction and Analysis of Systems},
pages = {337–340},
numpages = {4},
location = {Budapest, Hungary},
series = {TACAS'08/ETAPS'08}
}

@article{BlanchetCCFMMMR03,
author = {Blanchet, Bruno and Cousot, Patrick and Cousot, Radhia and Feret, J\'{e}rome and Mauborgne, Laurent and Min\'{e}, Antoine and Monniaux, David and Rival, Xavier},
title = {A static analyzer for large safety-critical software},
year = {2003},
issue_date = {May 2003},
publisher = {Association for Computing Machinery},
address = {New York, NY, USA},
volume = {38},
number = {5},
issn = {0362-1340},
url = {https://doi.org/10.1145/780822.781153},
doi = {10.1145/780822.781153},
journal = {SIGPLAN Not.},
month = may,
pages = {196–207},
numpages = {12}
}

@inproceedings{CousotC77,
author = {Cousot, Patrick and Cousot, Radhia},
title = {Abstract interpretation: a unified lattice model for static analysis of programs by construction or approximation of fixpoints},
year = {1977},
isbn = {9781450373500},
publisher = {Association for Computing Machinery},
address = {New York, NY, USA},
url = {https://doi.org/10.1145/512950.512973},
doi = {10.1145/512950.512973},
booktitle = {Proceedings of the 4th ACM SIGACT-SIGPLAN Symposium on Principles of Programming Languages},
pages = {238–252},
numpages = {15},
location = {Los Angeles, California},
series = {POPL '77}
}

@inproceedings{JhalaM06,
author = {Jhala, Ranjit and McMillan, K. L.},
title = {A practical and complete approach to predicate refinement},
year = {2006},
isbn = {3540330569},
publisher = {Springer-Verlag},
address = {Berlin, Heidelberg},
url = {https://doi.org/10.1007/11691372_33},
doi = {10.1007/11691372_33},
booktitle = {Proceedings of the 12th International Conference on Tools and Algorithms for the Construction and Analysis of Systems},
pages = {459–473},
numpages = {15},
location = {Vienna, Austria},
series = {TACAS'06}
}

@inproceedings{Mcmillan10,
author = {McMillan, Kenneth L.},
title = {Lazy annotation for program testing and verification},
year = {2010},
isbn = {364214294X},
publisher = {Springer-Verlag},
address = {Berlin, Heidelberg},
url = {https://doi.org/10.1007/978-3-642-14295-6_10},
doi = {10.1007/978-3-642-14295-6_10},
booktitle = {Proceedings of the 22nd International Conference on Computer Aided Verification},
pages = {104–118},
numpages = {15},
location = {Edinburgh, UK},
series = {CAV'10}
}

@inproceedings{ErnstCGK99,
author = {Ernst, Michael D. and Cockrell, Jake and Griswold, William G. and Notkin, David},
title = {Dynamically discovering likely program invariants to support program evolution},
year = {1999},
isbn = {1581130740},
publisher = {Association for Computing Machinery},
address = {New York, NY, USA},
url = {https://doi.org/10.1145/302405.302467},
doi = {10.1145/302405.302467},
booktitle = {Proceedings of the 21st International Conference on Software Engineering},
pages = {213–224},
numpages = {12},
location = {Los Angeles, California, USA},
series = {ICSE '99}
}

@article{PadhiSM16,
author = {Padhi, Saswat and Sharma, Rahul and Millstein, Todd},
title = {Data-driven precondition inference with learned features},
year = {2016},
issue_date = {June 2016},
publisher = {Association for Computing Machinery},
address = {New York, NY, USA},
volume = {51},
number = {6},
issn = {0362-1340},
url = {https://doi.org/10.1145/2980983.2908099},
doi = {10.1145/2980983.2908099},
journal = {SIGPLAN Not.},
month = jun,
pages = {42–56},
numpages = {15}
}

@misc{CLI2INV19,
      title={CLN2INV: Learning Loop Invariants with Continuous Logic Networks}, 
      author={Gabriel Ryan and Justin Wong and Jianan Yao and Ronghui Gu and Suman Jana},
      year={2019},
      eprint={1909.11542},
      archivePrefix={arXiv},
      primaryClass={cs.LG},
      url={https://arxiv.org/abs/1909.11542}, 
}

@inproceedings{YaoRWJG20,
author = {Yao, Jianan and Ryan, Gabriel and Wong, Justin and Jana, Suman and Gu, Ronghui},
title = {Learning nonlinear loop invariants with gated continuous logic networks},
year = {2020},
isbn = {9781450376136},
publisher = {Association for Computing Machinery},
address = {New York, NY, USA},
url = {https://doi.org/10.1145/3385412.3385986},
doi = {10.1145/3385412.3385986},
booktitle = {Proceedings of the 41st ACM SIGPLAN Conference on Programming Language Design and Implementation},
pages = {106–120},
numpages = {15},
location = {London, UK},
series = {PLDI 2020}
}

@inproceedings{YuWW23,
author = {Yu, Shiwen and Wang, Ting and Wang, Ji},
title = {Loop Invariant Inference through SMT Solving Enhanced Reinforcement Learning},
year = {2023},
isbn = {9798400702211},
publisher = {Association for Computing Machinery},
address = {New York, NY, USA},
url = {https://doi.org/10.1145/3597926.3598047},
doi = {10.1145/3597926.3598047},
booktitle = {Proceedings of the 32nd ACM SIGSOFT International Symposium on Software Testing and Analysis},
pages = {175–187},
numpages = {13},
location = {Seattle, WA, USA},
series = {ISSTA 2023}
}

@inproceedings{WenCSXQHLCT24,
author = {Wen, Cheng and Cao, Jialun and Su, Jie and Xu, Zhiwu and Qin, Shengchao and He, Mengda and Li, Haokun and Cheung, Shing-Chi and Tian, Cong},
title = {Enchanting Program Specification Synthesis by Large Language Models Using Static Analysis and Program Verification},
year = {2024},
isbn = {978-3-031-65629-3},
publisher = {Springer-Verlag},
address = {Berlin, Heidelberg},
url = {https://doi.org/10.1007/978-3-031-65630-9_16},
doi = {10.1007/978-3-031-65630-9_16},
booktitle = {Computer Aided Verification: 36th International Conference, CAV 2024, Montreal, QC, Canada, July 24–27, 2024, Proceedings, Part II},
pages = {302–328},
numpages = {27},
location = {Montreal, QC, Canada}
}

@inproceedings{
wu2024lemur,
title={Lemur: Integrating Large Language Models in Automated Program Verification},
author={Haoze Wu and Clark Barrett and Nina Narodytska},
booktitle={The Twelfth International Conference on Learning Representations},
year={2024},
url={https://openreview.net/forum?id=Q3YaCghZNt}
}

@article{Coq97,
author = {Barras, Bruno and Boutin, Samuel and Cornes, Cristina and Courant, Judicaël and Filliâtre, Jean-Christophe and Giménez, Eduardo and Herbelin, Hugo and Huet, Gérard and Muñoz, César and Murthy, Chetan and Parent-vigouroux, Catherine and Paulin-Mohring, Christine and Saïbi, Amokrane and Werner, Benjamin},
year = {1997},
month = {06},
pages = {},
title = {The Coq Proof Assistant Reference Manual : Version 6.1}
}

@book{Isabelle02,
author = {Nipkow, Tobias and Wenzel, Markus and Paulson, Lawrence C.},
title = {Isabelle/HOL: a proof assistant for higher-order logic},
year = {2002},
isbn = {3540433767},
publisher = {Springer-Verlag},
address = {Berlin, Heidelberg}
}

@InProceedings{Lean15,
author="de Moura, Leonardo
and Kong, Soonho
and Avigad, Jeremy
and van Doorn, Floris
and von Raumer, Jakob",
editor="Felty, Amy P.
and Middeldorp, Aart",
title="The Lean Theorem Prover (System Description)",
booktitle="Automated Deduction - CADE-25",
year="2015",
publisher="Springer International Publishing",
address="Cham",
pages="378--388",
isbn="978-3-319-21401-6"
}

@inproceedings{
jiang2023draft,
title={Draft, Sketch, and Prove: Guiding Formal Theorem Provers with Informal Proofs},
author={Albert Qiaochu Jiang and Sean Welleck and Jin Peng Zhou and Timothee Lacroix and Jiacheng Liu and Wenda Li and Mateja Jamnik and Guillaume Lample and Yuhuai Wu},
booktitle={The Eleventh International Conference on Learning Representations },
year={2023},
url={https://openreview.net/forum?id=SMa9EAovKMC}
}

@inproceedings{
zhou2024dont,
title={Don't Trust: Verify -- Grounding {LLM} Quantitative Reasoning with Autoformalization},
author={Jin Peng Zhou and Charles E Staats and Wenda Li and Christian Szegedy and Kilian Q Weinberger and Yuhuai Wu},
booktitle={The Twelfth International Conference on Learning Representations},
year={2024},
url={https://openreview.net/forum?id=V5tdi14ple}
}

@article{kamoi-etal-2024-llms,
    title = "When Can {LLM}s Actually Correct Their Own Mistakes? A Critical Survey of Self-Correction of {LLM}s",
    author = "Kamoi, Ryo  and
      Zhang, Yusen  and
      Zhang, Nan  and
      Han, Jiawei  and
      Zhang, Rui",
    journal = "Transactions of the Association for Computational Linguistics",
    volume = "12",
    year = "2024",
    address = "Cambridge, MA",
    publisher = "MIT Press",
    url = "https://aclanthology.org/2024.tacl-1.78/",
    doi = "10.1162/tacl_a_00713",
    pages = "1417--1440"
}

@inproceedings{WuJLRSJS22,
  author       = {Yuhuai Wu and
                  Albert Qiaochu Jiang and
                  Wenda Li and
                  Markus N. Rabe and
                  Charles Staats and
                  Mateja Jamnik and
                  Christian Szegedy},
  editor       = {Sanmi Koyejo and
                  S. Mohamed and
                  A. Agarwal and
                  Danielle Belgrave and
                  K. Cho and
                  A. Oh},
  title        = {Autoformalization with Large Language Models},
  booktitle    = {Advances in Neural Information Processing Systems 35: Annual Conference
                  on Neural Information Processing Systems 2022, NeurIPS 2022, New Orleans,
                  LA, USA, November 28 - December 9, 2022},
  year         = {2022},
  url          = {http://papers.nips.cc/paper\_files/paper/2022/hash/d0c6bc641a56bebee9d985b937307367-Abstract-Conference.html},
  bibsource    = {dblp computer science bibliography, https://dblp.org}
}

@article{Clause2Inv,
author = {Cao, Weining and Wu, Guangyuan and Xu, Tangzhi and Yao, Yuan and Wei, Hengfeng and Chen, Taolue and Ma, Xiaoxing},
title = {Clause2Inv: A Generate-Combine-Check Framework for Loop Invariant Inference},
year = {2025},
issue_date = {July 2025},
publisher = {Association for Computing Machinery},
address = {New York, NY, USA},
volume = {2},
number = {ISSTA},
url = {https://doi.org/10.1145/3728920},
doi = {10.1145/3728920},
journal = {Proc. ACM Softw. Eng.},
month = jun,
articleno = {ISSTA045},
numpages = {22}
}

@inproceedings{LiuWFCY24,
author = {Liu, Chang and Wu, Xiwei and Feng, Yuan and Cao, Qinxiang and Yan, Junchi},
title = {Towards general loop invariant generation: a benchmark of programs with memory manipulation},
year = {2024},
isbn = {9798331314385},
publisher = {Curran Associates Inc.},
address = {Red Hook, NY, USA},
booktitle = {Proceedings of the 38th International Conference on Neural Information Processing Systems},
articleno = {4101},
numpages = {26},
location = {Vancouver, BC, Canada},
series = {NIPS '24}
}

@InProceedings{cvc5,
author="Barbosa, Haniel
and Barrett, Clark
and Brain, Martin
and Kremer, Gereon
and Lachnitt, Hanna
and Mann, Makai
and Mohamed, Abdalrhman
and Mohamed, Mudathir
and Niemetz, Aina
and N{\"o}tzli, Andres
and Ozdemir, Alex
and Preiner, Mathias
and Reynolds, Andrew
and Sheng, Ying
and Tinelli, Cesare
and Zohar, Yoni",
editor="Fisman, Dana
and Rosu, Grigore",
title="cvc5: A Versatile and Industrial-Strength SMT Solver",
booktitle="Tools and Algorithms for the Construction and Analysis of Systems",
year="2022",
publisher="Springer International Publishing",
address="Cham",
pages="415--442",
isbn="978-3-030-99524-9"
}

@article{TaLKC19,
author = {Ta, Quang-Trung and Le, Ton Chanh and Khoo, Siau-Cheng and Chin, Wei-Ngan},
title = {Automated mutual induction proof in separation logic},
year = {2019},
issue_date = {Apr 2019},
publisher = {Springer-Verlag},
address = {Berlin, Heidelberg},
volume = {31},
number = {2},
issn = {0934-5043},
url = {https://doi.org/10.1007/s00165-018-0471-5},
doi = {10.1007/s00165-018-0471-5},
journal = {Form. Asp. Comput.},
month = apr,
pages = {207–230},
numpages = {24}
}

@article{antopoulosCounterexampleguidedApproachFinding2016,
  title = {A {{Counterexample-guided Approach}} to {{Finding Numerical Invariants}}},
  author = {Antopoulos, Timos and Ruef, Andrew and Hicks, Michael},
  date = {2016}
}

@online{ChatGPT4,
  title = {{{GPT-4 Technical Report}}},
  author = {OpenAI and Achiam, Josh and Adler, Steven and Agarwal, Sandhini and Ahmad, Lama and Akkaya, Ilge and Aleman, Florencia Leoni and Almeida, Diogo and Altenschmidt, Janko and Altman, Sam and Anadkat, Shyamal and Avila, Red and Babuschkin, Igor and Balaji, Suchir and Balcom, Valerie and Baltescu, Paul and Bao, Haiming and Bavarian, Mohammad and Belgum, Jeff and Bello, Irwan and Berdine, Jake and Bernadett-Shapiro, Gabriel and Berner, Christopher and Bogdonoff, Lenny and Boiko, Oleg and Boyd, Madelaine and Brakman, Anna-Luisa and Brockman, Greg and Brooks, Tim and Brundage, Miles and Button, Kevin and Cai, Trevor and Campbell, Rosie and Cann, Andrew and Carey, Brittany and Carlson, Chelsea and Carmichael, Rory and Chan, Brooke and Chang, Che and Chantzis, Fotis and Chen, Derek and Chen, Sully and Chen, Ruby and Chen, Jason and Chen, Mark and Chess, Ben and Cho, Chester and Chu, Casey and Chung, Hyung Won and Cummings, Dave and Currier, Jeremiah and Dai, Yunxing and Decareaux, Cory and Degry, Thomas and Deutsch, Noah and Deville, Damien and Dhar, Arka and Dohan, David and Dowling, Steve and Dunning, Sheila and Ecoffet, Adrien and Eleti, Atty and Eloundou, Tyna and Farhi, David and Fedus, Liam and Felix, Niko and Fishman, Simón Posada and Forte, Juston and Fulford, Isabella and Gao, Leo and Georges, Elie and Gibson, Christian and Goel, Vik and Gogineni, Tarun and Goh, Gabriel and Gontijo-Lopes, Rapha and Gordon, Jonathan and Grafstein, Morgan and Gray, Scott and Greene, Ryan and Gross, Joshua and Gu, Shixiang Shane and Guo, Yufei and Hallacy, Chris and Han, Jesse and Harris, Jeff and He, Yuchen and Heaton, Mike and Heidecke, Johannes and Hesse, Chris and Hickey, Alan and Hickey, Wade and Hoeschele, Peter and Houghton, Brandon and Hsu, Kenny and Hu, Shengli and Hu, Xin and Huizinga, Joost and Jain, Shantanu and Jain, Shawn and Jang, Joanne and Jiang, Angela and Jiang, Roger and Jin, Haozhun and Jin, Denny and Jomoto, Shino and Jonn, Billie and Jun, Heewoo and Kaftan, Tomer and Kaiser, Łukasz and Kamali, Ali and Kanitscheider, Ingmar and Keskar, Nitish Shirish and Khan, Tabarak and Kilpatrick, Logan and Kim, Jong Wook and Kim, Christina and Kim, Yongjik and Kirchner, Jan Hendrik and Kiros, Jamie and Knight, Matt and Kokotajlo, Daniel and Kondraciuk, Łukasz and Kondrich, Andrew and Konstantinidis, Aris and Kosic, Kyle and Krueger, Gretchen and Kuo, Vishal and Lampe, Michael and Lan, Ikai and Lee, Teddy and Leike, Jan and Leung, Jade and Levy, Daniel and Li, Chak Ming and Lim, Rachel and Lin, Molly and Lin, Stephanie and Litwin, Mateusz and Lopez, Theresa and Lowe, Ryan and Lue, Patricia and Makanju, Anna and Malfacini, Kim and Manning, Sam and Markov, Todor and Markovski, Yaniv and Martin, Bianca and Mayer, Katie and Mayne, Andrew and McGrew, Bob and McKinney, Scott Mayer and McLeavey, Christine and McMillan, Paul and McNeil, Jake and Medina, David and Mehta, Aalok and Menick, Jacob and Metz, Luke and Mishchenko, Andrey and Mishkin, Pamela and Monaco, Vinnie and Morikawa, Evan and Mossing, Daniel and Mu, Tong and Murati, Mira and Murk, Oleg and Mély, David and Nair, Ashvin and Nakano, Reiichiro and Nayak, Rajeev and Neelakantan, Arvind and Ngo, Richard and Noh, Hyeonwoo and Ouyang, Long and O'Keefe, Cullen and Pachocki, Jakub and Paino, Alex and Palermo, Joe and Pantuliano, Ashley and Parascandolo, Giambattista and Parish, Joel and Parparita, Emy and Passos, Alex and Pavlov, Mikhail and Peng, Andrew and Perelman, Adam and Peres, Filipe de Avila Belbute and Petrov, Michael and Pinto, Henrique Ponde de Oliveira and Michael and Pokorny and Pokrass, Michelle and Pong, Vitchyr H. and Powell, Tolly and Power, Alethea and Power, Boris and Proehl, Elizabeth and Puri, Raul and Radford, Alec and Rae, Jack and Ramesh, Aditya and Raymond, Cameron and Real, Francis and Rimbach, Kendra and Ross, Carl and Rotsted, Bob and Roussez, Henri and Ryder, Nick and Saltarelli, Mario and Sanders, Ted and Santurkar, Shibani and Sastry, Girish and Schmidt, Heather and Schnurr, David and Schulman, John and Selsam, Daniel and Sheppard, Kyla and Sherbakov, Toki and Shieh, Jessica and Shoker, Sarah and Shyam, Pranav and Sidor, Szymon and Sigler, Eric and Simens, Maddie and Sitkin, Jordan and Slama, Katarina and Sohl, Ian and Sokolowsky, Benjamin and Song, Yang and Staudacher, Natalie and Such, Felipe Petroski and Summers, Natalie and Sutskever, Ilya and Tang, Jie and Tezak, Nikolas and Thompson, Madeleine B. and Tillet, Phil and Tootoonchian, Amin and Tseng, Elizabeth and Tuggle, Preston and Turley, Nick and Tworek, Jerry and Uribe, Juan Felipe Cerón and Vallone, Andrea and Vijayvergiya, Arun and Voss, Chelsea and Wainwright, Carroll and Wang, Justin Jay and Wang, Alvin and Wang, Ben and Ward, Jonathan and Wei, Jason and Weinmann, C. J. and Welihinda, Akila and Welinder, Peter and Weng, Jiayi and Weng, Lilian and Wiethoff, Matt and Willner, Dave and Winter, Clemens and Wolrich, Samuel and Wong, Hannah and Workman, Lauren and Wu, Sherwin and Wu, Jeff and Wu, Michael and Xiao, Kai and Xu, Tao and Yoo, Sarah and Yu, Kevin and Yuan, Qiming and Zaremba, Wojciech and Zellers, Rowan and Zhang, Chong and Zhang, Marvin and Zhao, Shengjia and Zheng, Tianhao and Zhuang, Juntang and Zhuk, William and Zoph, Barret},
  date = {2024-03-04},
  eprint = {2303.08774},
  eprinttype = {arXiv},
  eprintclass = {cs},
  doi = {10.48550/arXiv.2303.08774},
  url = {http://arxiv.org/abs/2303.08774},
  urldate = {2025-05-06},
  pubstate = {prepublished}
}

@inproceedings{Code2Inv18,
  title = {Learning Loop Invariants for Program Verification},
  booktitle = {Advances in Neural Information Processing Systems 31: {{Annual}} Conference on Neural Information Processing Systems 2018, {{NeurIPS}} 2018, December 3-8, 2018, Montréal, Canada},
  author = {Si, Xujie and Dai, Hanjun and Raghothaman, Mukund and Naik, Mayur and Song, Le},
  editor = {Bengio, Samy and Wallach, Hanna M. and Larochelle, Hugo and Grauman, Kristen and Cesa-Bianchi, Nicolò and Garnett, Roman},
  date = {2018},
  pages = {7762--7773},
  url = {https://proceedings.neurips.cc/paper/2018/hash/65b1e92c585fd4c2159d5f33b5030ff2-Abstract.html},
  bibsource = {dblp computer science bibliography, https://dblp.org},
  eventtitle = {Advances in {{Neural Information Processing Systems}}}
}

@inproceedings{Code2Inv20,
  title = {{{Code2Inv}}: {{A}} Deep Learning Framework for Program Verification},
  booktitle = {Computer Aided Verification - 32nd International Conference, {{CAV}} 2020, Los Angeles, {{CA}}, {{USA}}, July 21-24, 2020, Proceedings, Part {{II}}},
  author = {Si, Xujie and Naik, Aaditya and Dai, Hanjun and Naik, Mayur and Song, Le},
  date = {2020},
  series = {Lecture Notes in Computer Science},
  volume = {12225},
  pages = {151--164},
  publisher = {Springer},
  doi = {10.1007/978-3-030-53291-8\_9},
  url = {https://doi.org/10.1007/978-3-030-53291-8_9},
  eventtitle = {International {{Conference}} on {{Computer Aided Verification}}}
}

@online{DeepSeekV3,
  title = {{{DeepSeek-V3 Technical Report}}},
  author = {DeepSeek-AI and Liu, Aixin and Feng, Bei and Xue, Bing and Wang, Bingxuan and Wu, Bochao and Lu, Chengda and Zhao, Chenggang and Deng, Chengqi and Zhang, Chenyu and Ruan, Chong and Dai, Damai and Guo, Daya and Yang, Dejian and Chen, Deli and Ji, Dongjie and Li, Erhang and Lin, Fangyun and Dai, Fucong and Luo, Fuli and Hao, Guangbo and Chen, Guanting and Li, Guowei and Zhang, H. and Bao, Han and Xu, Hanwei and Wang, Haocheng and Zhang, Haowei and Ding, Honghui and Xin, Huajian and Gao, Huazuo and Li, Hui and Qu, Hui and Cai, J. L. and Liang, Jian and Guo, Jianzhong and Ni, Jiaqi and Li, Jiashi and Wang, Jiawei and Chen, Jin and Chen, Jingchang and Yuan, Jingyang and Qiu, Junjie and Li, Junlong and Song, Junxiao and Dong, Kai and Hu, Kai and Gao, Kaige and Guan, Kang and Huang, Kexin and Yu, Kuai and Wang, Lean and Zhang, Lecong and Xu, Lei and Xia, Leyi and Zhao, Liang and Wang, Litong and Zhang, Liyue and Li, Meng and Wang, Miaojun and Zhang, Mingchuan and Zhang, Minghua and Tang, Minghui and Li, Mingming and Tian, Ning and Huang, Panpan and Wang, Peiyi and Zhang, Peng and Wang, Qiancheng and Zhu, Qihao and Chen, Qinyu and Du, Qiushi and Chen, R. J. and Jin, R. L. and Ge, Ruiqi and Zhang, Ruisong and Pan, Ruizhe and Wang, Runji and Xu, Runxin and Zhang, Ruoyu and Chen, Ruyi and Li, S. S. and Lu, Shanghao and Zhou, Shangyan and Chen, Shanhuang and Wu, Shaoqing and Ye, Shengfeng and Ye, Shengfeng and Ma, Shirong and Wang, Shiyu and Zhou, Shuang and Yu, Shuiping and Zhou, Shunfeng and Pan, Shuting and Wang, T. and Yun, Tao and Pei, Tian and Sun, Tianyu and Xiao, W. L. and Zeng, Wangding and Zhao, Wanjia and An, Wei and Liu, Wen and Liang, Wenfeng and Gao, Wenjun and Yu, Wenqin and Zhang, Wentao and Li, X. Q. and Jin, Xiangyue and Wang, Xianzu and Bi, Xiao and Liu, Xiaodong and Wang, Xiaohan and Shen, Xiaojin and Chen, Xiaokang and Zhang, Xiaokang and Chen, Xiaosha and Nie, Xiaotao and Sun, Xiaowen and Wang, Xiaoxiang and Cheng, Xin and Liu, Xin and Xie, Xin and Liu, Xingchao and Yu, Xingkai and Song, Xinnan and Shan, Xinxia and Zhou, Xinyi and Yang, Xinyu and Li, Xinyuan and Su, Xuecheng and Lin, Xuheng and Li, Y. K. and Wang, Y. Q. and Wei, Y. X. and Zhu, Y. X. and Zhang, Yang and Xu, Yanhong and Xu, Yanhong and Huang, Yanping and Li, Yao and Zhao, Yao and Sun, Yaofeng and Li, Yaohui and Wang, Yaohui and Yu, Yi and Zheng, Yi and Zhang, Yichao and Shi, Yifan and Xiong, Yiliang and He, Ying and Tang, Ying and Piao, Yishi and Wang, Yisong and Tan, Yixuan and Ma, Yiyang and Liu, Yiyuan and Guo, Yongqiang and Wu, Yu and Ou, Yuan and Zhu, Yuchen and Wang, Yuduan and Gong, Yue and Zou, Yuheng and He, Yujia and Zha, Yukun and Xiong, Yunfan and Ma, Yunxian and Yan, Yuting and Luo, Yuxiang and You, Yuxiang and Liu, Yuxuan and Zhou, Yuyang and Wu, Z. F. and Ren, Z. Z. and Ren, Zehui and Sha, Zhangli and Fu, Zhe and Xu, Zhean and Huang, Zhen and Zhang, Zhen and Xie, Zhenda and Zhang, Zhengyan and Hao, Zhewen and Gou, Zhibin and Ma, Zhicheng and Yan, Zhigang and Shao, Zhihong and Xu, Zhipeng and Wu, Zhiyu and Zhang, Zhongyu and Li, Zhuoshu and Gu, Zihui and Zhu, Zijia and Liu, Zijun and Li, Zilin and Xie, Ziwei and Song, Ziyang and Gao, Ziyi and Pan, Zizheng},
  date = {2025-02-18},
  eprint = {2412.19437},
  eprinttype = {arXiv},
  eprintclass = {cs},
  doi = {10.48550/arXiv.2412.19437},
  url = {http://arxiv.org/abs/2412.19437},
  urldate = {2025-05-06},
  pubstate = {prepublished}
}

@misc{DeepSeekR1,
      title={DeepSeek-R1: Incentivizing Reasoning Capability in LLMs via Reinforcement Learning}, 
      author={DeepSeek-AI and Daya Guo and Dejian Yang and Haowei Zhang and Junxiao Song and Ruoyu Zhang and Runxin Xu and Qihao Zhu and Shirong Ma and Peiyi Wang and Xiao Bi and Xiaokang Zhang and Xingkai Yu and Yu Wu and Z. F. Wu and Zhibin Gou and Zhihong Shao and Zhuoshu Li and Ziyi Gao and Aixin Liu and Bing Xue and Bingxuan Wang and Bochao Wu and Bei Feng and Chengda Lu and Chenggang Zhao and Chengqi Deng and Chenyu Zhang and Chong Ruan and Damai Dai and Deli Chen and Dongjie Ji and Erhang Li and Fangyun Lin and Fucong Dai and Fuli Luo and Guangbo Hao and Guanting Chen and Guowei Li and H. Zhang and Han Bao and Hanwei Xu and Haocheng Wang and Honghui Ding and Huajian Xin and Huazuo Gao and Hui Qu and Hui Li and Jianzhong Guo and Jiashi Li and Jiawei Wang and Jingchang Chen and Jingyang Yuan and Junjie Qiu and Junlong Li and J. L. Cai and Jiaqi Ni and Jian Liang and Jin Chen and Kai Dong and Kai Hu and Kaige Gao and Kang Guan and Kexin Huang and Kuai Yu and Lean Wang and Lecong Zhang and Liang Zhao and Litong Wang and Liyue Zhang and Lei Xu and Leyi Xia and Mingchuan Zhang and Minghua Zhang and Minghui Tang and Meng Li and Miaojun Wang and Mingming Li and Ning Tian and Panpan Huang and Peng Zhang and Qiancheng Wang and Qinyu Chen and Qiushi Du and Ruiqi Ge and Ruisong Zhang and Ruizhe Pan and Runji Wang and R. J. Chen and R. L. Jin and Ruyi Chen and Shanghao Lu and Shangyan Zhou and Shanhuang Chen and Shengfeng Ye and Shiyu Wang and Shuiping Yu and Shunfeng Zhou and Shuting Pan and S. S. Li and Shuang Zhou and Shaoqing Wu and Shengfeng Ye and Tao Yun and Tian Pei and Tianyu Sun and T. Wang and Wangding Zeng and Wanjia Zhao and Wen Liu and Wenfeng Liang and Wenjun Gao and Wenqin Yu and Wentao Zhang and W. L. Xiao and Wei An and Xiaodong Liu and Xiaohan Wang and Xiaokang Chen and Xiaotao Nie and Xin Cheng and Xin Liu and Xin Xie and Xingchao Liu and Xinyu Yang and Xinyuan Li and Xuecheng Su and Xuheng Lin and X. Q. Li and Xiangyue Jin and Xiaojin Shen and Xiaosha Chen and Xiaowen Sun and Xiaoxiang Wang and Xinnan Song and Xinyi Zhou and Xianzu Wang and Xinxia Shan and Y. K. Li and Y. Q. Wang and Y. X. Wei and Yang Zhang and Yanhong Xu and Yao Li and Yao Zhao and Yaofeng Sun and Yaohui Wang and Yi Yu and Yichao Zhang and Yifan Shi and Yiliang Xiong and Ying He and Yishi Piao and Yisong Wang and Yixuan Tan and Yiyang Ma and Yiyuan Liu and Yongqiang Guo and Yuan Ou and Yuduan Wang and Yue Gong and Yuheng Zou and Yujia He and Yunfan Xiong and Yuxiang Luo and Yuxiang You and Yuxuan Liu and Yuyang Zhou and Y. X. Zhu and Yanhong Xu and Yanping Huang and Yaohui Li and Yi Zheng and Yuchen Zhu and Yunxian Ma and Ying Tang and Yukun Zha and Yuting Yan and Z. Z. Ren and Zehui Ren and Zhangli Sha and Zhe Fu and Zhean Xu and Zhenda Xie and Zhengyan Zhang and Zhewen Hao and Zhicheng Ma and Zhigang Yan and Zhiyu Wu and Zihui Gu and Zijia Zhu and Zijun Liu and Zilin Li and Ziwei Xie and Ziyang Song and Zizheng Pan and Zhen Huang and Zhipeng Xu and Zhongyu Zhang and Zhen Zhang},
      year={2025},
      eprint={2501.12948},
      archivePrefix={arXiv},
      primaryClass={cs.CL},
      url={https://arxiv.org/abs/2501.12948}, 
}

@misc{surveylogicerror,
      title={Empowering LLMs with Logical Reasoning: A Comprehensive Survey}, 
      author={Fengxiang Cheng and Haoxuan Li and Fenrong Liu and Robert van Rooij and Kun Zhang and Zhouchen Lin},
      year={2025},
      eprint={2502.15652},
      archivePrefix={arXiv},
      primaryClass={cs.AI},
      url={https://arxiv.org/abs/2502.15652}, 
}
% x1-LLM-loop is the bib file of Zhenyu

%% Appendix
%\newpage
%\appendix
%  \input{appendix}

\end{document}